\documentclass{article}
\usepackage{amsmath}
\usepackage{amssymb}
\usepackage[cmtip,arrow]{xy}
\usepackage{pb-diagram,pb-xy}

\begin{document}

\title{\textbf{Dirac approach to constrained submanifolds in a double loop
group: from WZNW to Poisson-Lie }$\sigma $\textbf{-model}\\
\smallskip\ }
\author{\textbf{H. Montani$^{\mathtt{a}}${\thanks{%
e-mail: \textit{\ hmontani@uaco.unpa.edu.ar }} } \& M. Zuccalli$^{\mathtt{b}}
${\thanks{%
e-mail: \textit{\ marcezuccalli@gmail.com }}}\textbf{\ } } \and \medskip \\
%EndAName
\textbf{$^{\mathtt{a}}$}\ \textit{CONICET \& Departamento de Ciencias
Exactas y Naturales, }\\
\textit{\ Universidad Nacional de la Patagonia Austral, }\\
\textit{\ (9011)Caleta Olivia, Argentina. }\\
\\
\textbf{$^{\mathtt{b}}$}\ \textit{\ Departamento de Matem\'{a}tica,
Universidad Nacional de La Plata,} \and \textit{Calle 50 esq. 115 (1900) La
Plata, Argentina}}
\maketitle

\begin{abstract}
We study the restriction to a family of second class constrained
submanifolds in the cotangent bundle of a double Lie group, equipped with a
2-cocycle extended symplectic form, building the corresponding Dirac
brackets. It is shown that, for a 2-cocycle vanishing on each isotropic
subspaces of the associated Manin triple, the Dirac bracket contains no
traces of the cocycle. We also investigate the restriction of the left
translation action of the double Lie group on its cotangent bundle, where it
fails in to be a symmetry a canonical transformation. However, the
hamiltonian symmetry is restored on some special submanifolds. The main
application is on loop groups, showing that a WZNW-type model on the double
Lie group with a quadratic Hamilton function in the momentum maps associated
with the left translation action on the cotangent bundle with the canonical
symplectic form, restricts to a collective system on some special
submanifolds. There, the lagrangian version coincides with so called
Poisson-Lie $\sigma $-model.
\end{abstract}

%\tableofcontents

\newpage Many relevant physical systems are modeled on Lie groups, taking
the corresponding cotangent bundles as their phase spaces. Among the finite
dimensional examples are the rigid bodies and its generalizations \cite%
{Arnold} and, in infinite dimension, the sigma and WZNW models are
outstanding field theories. Cotangent bundles are canonically symplectic
manifolds and they have rich structures underlying symmetries which involves
their Lie algebras and its dual spaces \cite{Abr-Mars},\cite{Arn-Given},\cite%
{Mars-Ratiu},\cite{Guillemin-Sternberg}. In some special situations, these
structures straightforwardly leads to integrability \cite{RSTS 0}.

In the last decades, the dynamic of integrable systems becomes more involved
with Lie groups: it turns out that most of them are deeply related to \emph{%
Poisson-Lie groups}, that is, Lie groups supplied with a compatible Poisson
structure \cite{Drinfeld}. A Poisson-Lie group has naturally associated a 
\emph{dual }Poisson-Lie group, and the \emph{double Lie group} build with
this dual pair is, in some sense, a self dual structure with a lot of nice
properties enriching the framework of integrability \cite{Drinfeld},\cite%
{STS 0},\cite{Lu-We}.

In reference \cite{CapMon JMP}, by regarding the cotangent bundle of this
kind of double Lie group as a fibration on one of its factors, the Dirac
method \cite{Dirac} was developed for dealing with the restriction to the
fibers of a dynamical system on whole space. In fact, these fibers turns to
be symplectic submanifolds of the contangent bundle of the double Lie group
equipped with the canonical symplectic form, turning the restriction to them
in a second class constraint problem.

Cotangent bundles of Lie groups equipped with the \emph{canonical symplectic
structure} are not enough to encode all the plethora of systems modelled on
Lie groups. For instance, the phase space of a sigma model with target space
the group manifold $G$ is the cotangent bundle $T^{\ast }LG$ of the loop
group $LG$ with the canonical symplectic form $\omega _{\mathtt{o}}$, and
the dynamics is determined by the election of the Hamilton function. On the
other side, the WZNW model shares the same configuration space, but it can
not be obtained from this phase space: no Hamilton function can be found
driving to Hamilton equations equivalent to the WZNW ones. In fact, it was
shown in ref. \cite{Fad-Takh} that the addition of Wess-Zumino term, the 
\emph{topological} term, to the action of the sigma model amounts to a
modification of the canonical Poisson brackets. It symplectic counterpart is
exhaustively studied in references \cite{Harnad}, where a cocycle extension
of the canonical symplectic form $\omega _{\mathtt{o}}$ is considered in
combination with the Marsden-Weinstein reduction by stages procedure in
order to recover the WZNW equation of motion \cite{MW}.

In this work we adapt the scheme developed in \cite{CapMon JMP} to the case
where the initial phase space is the cotangent bundle $T^{\ast }G$ of a
double Poisson-Lie group $G=G_{+}G_{-}$ supplied with the symplectic form $%
\omega _{c}$ obtained by modifying the canonical one by adding a 2-cocycle $%
c:\mathfrak{g}\otimes \mathfrak{g}\longrightarrow \mathbb{R}$ on the Lie
algebra $\mathfrak{g}$ of the double Lie group $G$. So, Hamiltonian systems
on this symplectic manifold are of the WZNW-type in the sense that their
Lagrangian counterpart exhibits the topological WZ-term. We consider
dynamical systems on the fiber of the fibration $T^{\ast }G\longrightarrow
T^{\ast }G_{-}$ which are symplectic submanifolds of $\left( T^{\ast
}G,\omega _{c}\right) $, so they can be addressed via the Dirac method for
second class constraints in analogous way as in \cite{CapMon JMP}. Thus, we
built the Dirac brackets to describe dynamical systems on these constrained
submanifolds in terms of the algebra of function on the whole space $T^{\ast
}G$. We investigate how the left action of the group $G$ on itself, lifted
to the cotangent bundle, restricts to the fibers and becomes in a symmetry
of these phase subspaces for some special fibers. These scheme comes to be
very useful when applied to the specific case of loop groups, where
amazingly the Dirac brackets lose the cocycle contributions. We work out the
restriction of a quadratic hamiltonian on $T^{\ast }G$ which becomes
collective on the same fibers where the left translation action turns to be
a symmetry, recovering a Poisson-Lie $\sigma $-model on each of these
fibers. All these facts bring the subject into the realm of Poisson-Lie
T-duality \cite{KS-1}, turning this machinery very useful for working on the
hamiltonian approach to it \cite{CM},\cite{CMZ}.

This work is organized as follows: we divided it in two main parts. In Part
I we concentrate in developing the Dirac machinery and symmetry issues for
the $2$-cocycle extended symplectic form on $T^{\ast }G$. Thus, in Section 1
we describe the phase space on a double Poisson-Lie Lie group and build the
fibration of constrained submanifolds. In the Section 2 we adapt the scheme
developed in \cite{CapMon JMP} to the case where the canonical symplectic
form is modified by adding a 2-cocycle. Section 3 is devoted to specialize
the construction to loop groups. In Section 4 we discuss the left
translation symmetry including the action by left translation of the
centrally extended group. Section 5 introduces the Hamilton equations for
the whole and the constrained spaces, and describes some properties of the
collective dynamics. In Part II we develop the main application of the
results of the Part I, mainly addressed to the loop group context. So, in
Section 6 we address a hamiltonian model which restrict to a collective one.
In Sections 7 and 8 we retrieve the Lagrange equations, introducing
explicitly the loop groups stage, making contact with the so called
Poisson-Lie $\sigma $-models. Finally, in the last Section some conclusions
are summarized.

\bigskip

\bigskip

\part{Phase spaces on double Lie groups and constrained systems}

In this first part we study the fibration $\Psi :G\times \mathfrak{g}^{\ast
}\longrightarrow G_{-}\times \mathfrak{g}_{-}^{\ast }$, for a double Lie
group $G=G_{+}G_{-}$, as the phase space of systems constrained to the
fibers $\Psi ^{-1}\left( g_{-},\eta _{-}\right) $. We adapt the Dirac's
machinery developed in $\cite{CapMon JMP}$ to the framework of $G\times 
\mathfrak{g}^{\ast }$ equipped with the $2$-cocycle extended symplectic
form, pointing to the loop groups stage. We also address the left
translation action of $G$ on $G\times \mathfrak{g}^{\ast }$ and its
restriction to $\Psi ^{-1}\left( g_{-},\eta _{-}\right) $, finding out the
fibers on which it turns a phase space symmetry with Ad-equivariant momentum
maps. Collective dynamics is then possible on some fibers, so we study its
properties in the current framework.

\section{The fibration $G\times \mathfrak{g}^{\ast }\longrightarrow
G_{-}\times \mathfrak{g}_{-}^{\ast }$}

Let us describe the framework for the main developments in this work (we
follow the definitions and notations in \cite{Lu-We}). Let $\left( \mathfrak{%
g},\mathfrak{g}_{+},\mathfrak{g}_{-}\right) $ be \emph{Manin triple}, that
means, $\mathfrak{g},\mathfrak{g}_{+},\mathfrak{g}_{-}$ are Lie algebras
such that $\mathfrak{g}=\mathfrak{g}_{+}\oplus \mathfrak{g}_{-}$ as a vector
space, and $\mathfrak{g}$ is supplied with a $\mathrm{ad}$-invariant
nondegenerate symmetric bilinear form $\left( ,\right) _{\mathfrak{g}}$
turning $\mathfrak{g}_{+},\mathfrak{g}_{-}$ into isotropic subspaces. Thus, $%
\mathfrak{g}$ is a \emph{double Lie algebra }and $\mathfrak{g}_{+},\mathfrak{%
g}_{-}$ are \emph{Lie bialgebras}, a couple of dual Lie bialgebras. The
associated connected simply connected Lie group $G,G_{+},G_{-}$ form a \emph{%
double Lie group }$G=G_{+}G_{-}$, with $G_{+}$ and $G_{-}$ being \emph{%
Poisson-Lie groups. }Let us denote the corresponding projectors $\Pi
_{G_{\pm }}:G\longrightarrow G_{\pm }$, $\Pi _{\mathfrak{g}_{\pm }}:%
\mathfrak{g}\longrightarrow \mathfrak{g}_{\pm }$ and $\Pi _{\mathfrak{g}%
_{\pm }^{\ast }}:\mathfrak{g}^{\ast }\longrightarrow \mathfrak{g}_{\pm
}^{\ast }$, that for short we frequently denote as $g_{\pm }=\Pi _{G_{\pm
}}g $, $X_{\pm }=\Pi _{\mathfrak{g}_{\pm }}X$ and $\eta _{\pm }=\Pi _{%
\mathfrak{g}_{\pm }^{\ast }}\eta $. The factorization of the elements like $%
g_{-}g_{+}\in G$ is denoted as 
\begin{equation*}
\begin{array}{ccc}
g_{+}^{h_{-}}:=\Pi _{G_{+}}\left( h_{-}g_{+}\right) & , & g_{-}^{h_{+}}:=\Pi
_{G_{-}}\left( h_{-}g_{+}\right)%
\end{array}%
\end{equation*}%
Indeed, the maps 
\begin{equation*}
G_{\mp }\times G_{\pm }\longrightarrow G_{\pm }\qquad /\qquad \left( h_{\mp
},g_{\pm }\right) \mapsto \Pi _{G_{\pm }}\left( h_{\mp }g_{\pm }\right)
=g_{\pm }^{h_{\mp }}
\end{equation*}%
amounts to be crossed actions between the factors, the so called \emph{%
dressing actions }\cite{STS 0},\cite{Lu-We}. The infinitesimal generator of
the dressing action of $G_{-}$ on $G_{+}$ at the point $g_{+}\in G_{+}$
gives rise to the antihomomorphism of Lie algebras $X_{-}\in \mathfrak{g}%
_{-}\mapsto g_{+}^{X_{-}}\in T_{g_{+}}G_{+}$, such that, for $X_{-},Y_{-}\in 
\mathfrak{g}_{-}$, $\left[ g_{+}^{X_{-}},g_{+}^{Y_{-}}\right] =-g_{+}^{\left[
X_{-},Y_{-}\right] _{\mathfrak{g}_{-}}}$.

Let $\psi $ be the identification $\mathfrak{g}\longrightarrow \mathfrak{g}%
^{\ast }$ provided by the nondegenerate bilinear form, and $\bar{\psi}$
denote its inverse, then the crossed adjoint actions are 
\begin{equation*}
\left\{ 
\begin{array}{l}
\mathrm{Ad}_{h_{+}^{-1}}^{G}X_{-}=h_{+}^{-1}h_{+}^{X_{-}}+\bar{\psi}\left(
Ad_{h_{+}^{-1}}^{\ast }\psi \left( X_{-}\right) \right) \\ 
\\ 
\mathrm{Ad}_{h_{-}}^{G}X_{+}=h_{-}^{X_{+}}h_{-}^{-1}+\bar{\psi}\left(
Ad_{h_{-}^{-1}}^{\ast }\psi \left( X_{+}\right) \right)%
\end{array}%
\right.
\end{equation*}%
where $Ad^{\ast }$ stands for the dual of the adjoint action of each group
factor on the dual of its Lie algebra (it relates with the coadjoint action $%
Ad^{\#}$as $Ad_{h_{+}}^{\#}:=Ad_{h_{+}^{-1}}^{\ast }$). This expression
allows to write the infinitesimal generators as $g_{+}^{X_{-}}=g_{+}\left(
\Pi _{\mathfrak{g}_{+}}\mathrm{Ad}_{g_{+}^{-1}}^{G}X_{-}\right) $ and $%
g_{-}^{X_{+}}=\left( \Pi _{\mathfrak{g}_{-}}\mathrm{Ad}_{g_{-}}^{G}X_{+}%
\right) g_{-}$.

The starting point of our developments is the cotangent bundle of the double
Lie group $G=G_{+}G_{-}$, realized as $G\times \mathfrak{g}^{\ast }$ by
using the left translation isomorphism, and the fibration defined by the
surjective submersion 
\begin{eqnarray*}
\Psi :G\times \mathfrak{g}^{\ast } &\longrightarrow &G_{-}\times \mathfrak{g}%
_{-}^{\ast } \\
\left( g,\eta \right) &\longmapsto &\left( g_{-},\eta _{-}\right)
\end{eqnarray*}

Let us name $\mathcal{N}\left( g_{-},\eta _{-}\right) $ the fiber on $\left(
g_{-},\eta _{-}\right) $, then%
\begin{equation*}
\mathcal{N}\left( g_{-},\eta _{-}\right) :=\Psi ^{-1}\left( g_{-},\eta
_{-}\right) =\left\{ \left( g_{+}g_{-},\eta _{+}+\eta _{-}\right) /g_{+}\in
G_{+}~,~\eta _{+}\in \mathfrak{g}_{+}^{\ast }\right\}
\end{equation*}

The differential $\Psi _{\ast }$ of the map $\Psi :G\times \mathfrak{g}%
^{\ast }\longrightarrow G_{-}\times \mathfrak{g}_{-}^{\ast }$ can be
obtained from%
\begin{equation*}
g^{-1}\dot{g}=Ad_{g_{-}^{-1}}^{G}g_{+}^{-1}\dot{g}_{+}+g_{-}^{-1}\dot{g}_{-}
\end{equation*}%
then, 
\begin{equation*}
\left. \Psi _{\ast }\left( gX,\xi \right) \right\vert _{\left( g,\eta
\right) }=\left( \left( \Pi _{\mathfrak{g}_{-}}\mathrm{Ad}%
_{g_{-}}^{G}X_{+}\right) g_{-}+g_{-}X_{-},\xi _{-}\right) _{\left(
g_{-},\eta _{-}\right) }
\end{equation*}%
The kernel of $\Psi _{\ast }$ coincides with $T\mathcal{N}\left( g_{-},\eta
_{-}\right) $, and it can be explicitly described as 
\begin{equation*}
\left. \ker \Psi _{\ast }\right\vert _{\left( g,\eta \right) }=\left\{
\left( g_{+}\left( \bar{\psi}\left( Ad_{g_{-}^{-1}}^{\ast }\psi \left(
X_{+}\right) \right) \right) g_{-},\xi _{+}\right) ~/~\left( X_{+},\xi
_{+}\right) \in \mathfrak{g}_{+}\oplus \mathfrak{g}_{+}^{\ast }\right\}
\end{equation*}%
The Dirac method is build from the annihilator of $\left. \ker \Psi _{\ast
}\right\vert _{\left( g,\eta \right) }$, that is the pullback of the
cotangent bundle of the base space $G_{-}\times \mathfrak{g}_{-}^{\ast }$.

\section{Centrally extended symplectic structures and the Dirac method}

In this section we adapt the Dirac bracket construction of ref. \cite{CapMon
JMP} to the case where canonical symplectic form on $G\times \mathfrak{g}%
^{\ast }$ is modified by adding a $\mathbb{R}$-valued $2$-cocycle on $%
\mathfrak{g}$ and we write the Dirac bracket on the submanifolds $\mathcal{N}%
_{c}\left( g_{-},\eta _{-}\right) $.

Let $C:G\longrightarrow \mathfrak{g}^{\ast }$ be a coadjoint $1$-cocycle,
that is, for $g,h\in G$ it satisfy 
\begin{equation*}
\begin{array}{ccc}
C\left( gh\right) =\mathrm{Ad}_{g^{-1}}^{G\ast }C\left( h\right) +C\left(
g\right) & , & \forall g,h\in G%
\end{array}%
\end{equation*}%
By considering $\hat{c}=-\left. dC\right\vert _{e}:\mathfrak{g}%
\longrightarrow \mathfrak{g}^{\ast }$, the 1-cocycle $C$ defines the
application $c:\mathfrak{g}\otimes \mathfrak{g}\longrightarrow \mathbb{R}$
given by 
\begin{equation*}
c\left( X,Y\right) :=\left\langle \hat{c}\left( X\right) ,Y\right\rangle
\end{equation*}%
Its easy to see that $c$ is bilinear, antisymmetric and verifies the Jacobi
identity. Then, $c$ is a $\mathbb{R}$-valued 2-cocycle on $\mathfrak{g}$
that satisfy the following condition

\begin{equation*}
c\left( Ad_{g}X,Ad_{g}Y\right) =c\left( X,Y\right) +\left\langle C\left(
g^{-1}\right) ,\left[ X,Y\right] \right\rangle
\end{equation*}%
So, let us consider the canonical symplectic form $\omega _{\mathtt{o}}$ in $%
T^{\ast }G$. Then, by adding $c$ one defines a new symplectic form on $%
T^{\ast }G$ given by%
\begin{eqnarray}
&&\langle \omega _{c},(v,\rho )\otimes (w,\xi )\rangle _{\left( g,\eta
\right) }  \label{cent ext symp form} \\
&:&=-\langle \rho ,g^{-1}w\rangle +\langle \xi ,g^{-1}v\rangle +\left\langle
\eta ,[g^{-1}v,g^{-1}w]\right\rangle +c\left( vg^{-1},wg^{-1}\right)  \notag
\end{eqnarray}%
for $(v,\rho ),(w,\xi )\in T_{\left( g,\eta \right) }^{\ast }G$.

The hamiltonian vector field of a function $\mathcal{F}:G\times \mathfrak{g}%
^{\ast }\longrightarrow \mathbb{R}$ is 
\begin{equation}
V_{\mathcal{F}}\left( g,\eta \right) =\left( g\delta \mathcal{F},ad_{\delta 
\mathcal{F}}^{\ast }\eta -g\mathbf{d}\mathcal{F}+\mathrm{Ad}_{g}^{G\ast }%
\hat{c}\left( Ad_{g}^{G}\delta \mathcal{F}\right) \right)
\label{ham vect cent exten sympl form}
\end{equation}%
for\textit{\ }$\left( g,\eta \right) \in $ $G\times \mathfrak{g}^{\ast }$,
and the associated Poisson bracket is 
\begin{equation*}
\left\{ \mathcal{F},\mathcal{G}\right\} _{c}\left( g,\eta \right)
=\left\langle \delta \mathcal{G},g\mathbf{d}\mathcal{F}\right\rangle
-\left\langle \mathbf{d}\mathcal{G},g\delta \mathcal{F}\right\rangle
-\left\langle \eta +C\left( g^{-1}\right) ,[\delta \mathcal{F},\delta 
\mathcal{G}]\right\rangle -c\left( \delta \mathcal{F},\delta \mathcal{G}%
\right)
\end{equation*}%
for $\mathcal{F},\mathcal{G}\in C^{\infty }(G\times \mathfrak{g}^{\ast })$.

Let us now proceed to adapt the Dirac method to the restrictions to the
fibers $\mathcal{N}\left( g_{-},\eta _{-}\right) $ when the centrally
extended symplectic form on $T^{\ast }G$ is considered.

\begin{description}
\item[Proposition:] $\left( \mathcal{N}\left( g_{-},\eta _{-}\right) ,\tilde{%
\omega}_{c}\right) $\textit{, where }$\tilde{\omega}_{c}$ \textit{is the
restriction to }$\mathcal{N}\left( g_{-},\eta _{-}\right) $\textit{\ of the
centrally extended canonical symplectic form }$\omega _{c}$ \textit{on }$%
G\times \mathfrak{g}^{\ast }$\textit{, is a symplectic manifold.}
\end{description}

\textbf{Proof: }Let us observe that the restriction of $\omega _{c}$ to the
kernel of $\Psi _{\ast }$, is given by

\begin{eqnarray*}
&&\left\langle \omega _{c},\left( g_{+}\left( \bar{\psi}\left(
Ad_{g_{-}^{-1}}^{\ast }\psi \left( X_{+}\right) \right) \right) g_{-},\xi
_{+}\right) \otimes \left( g_{+}\left( \bar{\psi}\left(
Ad_{g_{-}^{-1}}^{\ast }\psi \left( Y_{+}\right) \right) \right)
g_{-},\lambda _{+}\right) \right\rangle _{\left( g,\eta \right) } \\
&=&\left\langle \omega ,\left( g_{+}\left( \bar{\psi}\left(
Ad_{g_{-}^{-1}}^{\ast }\psi \left( X_{+}\right) \right) \right) g_{-},\xi
_{+}\right) \otimes \left( g_{+}\left( \bar{\psi}\left(
Ad_{g_{-}^{-1}}^{\ast }\psi \left( Y_{+}\right) \right) \right)
g_{-},\lambda _{+}\right) \right\rangle _{\left( g,\eta \right) } \\
&&+c\left( g_{+}\left( \bar{\psi}\left( Ad_{g_{-}^{-1}}^{\ast }\psi \left(
X_{+}\right) \right) \right) g_{-},g_{+}\left( \bar{\psi}\left(
Ad_{g_{-}^{-1}}^{\ast }\psi \left( Y_{+}\right) \right) \right) g_{-}\right)
\end{eqnarray*}%
for $\left( g_{+}\left( \bar{\psi}\left( Ad_{g_{-}^{-1}}^{\ast }\psi \left(
X_{+}\right) \right) \right) g_{-},\xi _{+}\right) ,\left( g_{+}\left( \bar{%
\psi}\left( Ad_{g_{-}^{-1}}^{\ast }\psi \left( Y_{+}\right) \right) \right)
g_{-},\lambda _{+}\right) \in \left. \ker \Psi _{\ast }\right\vert _{\left(
g,\eta \right) }$. Then, is easy to see that there are not null vectors of $%
\omega _{c}$ on $T\Psi ^{-1}\left( g_{-},\eta _{-}\right) =\left. \ker \Psi
_{\ast }\right\vert _{\left( g,\eta \right) }$.$\blacksquare $

\begin{description}
\item[Corollary:] $\left( T_{\left( g,\eta \right) }\mathcal{N}\left(
g_{-},\eta _{-}\right) \right) ^{\bot \omega _{c}}\cap T_{\left( g,\eta
\right) }\mathcal{N}\left( g_{-},\eta _{-}\right) =\left\{ 0\right\} $%
\textit{, then }$\Psi ^{-1}\left( g_{-},\eta _{-}\right) $ \textit{is a }%
\emph{second class constraint.}
\end{description}

Here, $\left( T_{\left( g,\eta \right) }\mathcal{N}\left( g_{-},\eta
_{-}\right) \right) ^{\bot \omega _{c}}$ denotes the orthogonal complement
in relation with the symplectic structure $\omega _{c}$.

Now, we choose a basis for $T_{\left( g_{-},\eta _{-}\right) }^{\ast }\left(
G_{-}\times \mathfrak{g}_{-}^{\ast }\right) \cong \mathfrak{g}_{-}^{\ast
}\oplus \mathfrak{g}_{-}$, where we used the left trivialization of $%
T_{\left( g_{-},\eta _{-}\right) }^{\ast }G_{-}$. The basis $\left\{
T_{a}\right\} $ of $\mathfrak{g}_{+}$ and the basis $\left\{ T^{a}\right\} $
of $\mathfrak{g}_{-}$ provides a set of linearly independent 1-forms on $%
G_{-}\times \mathfrak{g}_{-}^{\ast }$: 
\begin{eqnarray*}
\alpha _{a} &=&\left( L_{g_{-}^{-1}}^{\ast }\psi \left( T_{a}\right)
,0\right) \in T_{\left( g_{-},\eta _{-}\right) }^{\ast }G_{-}\times 
\mathfrak{g}_{-}^{\ast } \\
\beta _{a} &=&\left( 0,T^{a}\right) \in T_{\left( g_{-},\eta _{-}\right)
}^{\ast }G_{-}\times \mathfrak{g}_{-}^{\ast }
\end{eqnarray*}%
such that, for any $\left( v_{-},\xi _{-}\right) \in T_{\left( g_{-},\eta
_{-}\right) }G_{-}\times \mathfrak{g}_{-}^{\ast }$%
\begin{eqnarray*}
\left\langle \alpha _{a},\left( v_{-},\xi _{-}\right) \right\rangle _{\left(
g_{-},\eta _{-}\right) } &=&\left( T_{a},g_{-}^{-1}v_{-}\right) _{\mathfrak{g%
}} \\
\left\langle \beta _{a},\left( v_{-},\xi _{-}\right) \right\rangle _{\left(
g_{-},\eta _{-}\right) } &=&\left\langle \xi _{-},T^{a}\right\rangle
\end{eqnarray*}%
The set of pullbacks $\left\{ \Psi ^{\ast }\alpha _{a},\Psi ^{\ast }\beta
^{a}\right\} _{a}$ is linearly independent, with $T\mathcal{N}\left(
g_{-},\eta _{-}\right) $ being its null distribution.

The associated hamiltonian vector fields $V_{\Psi ^{\ast }\alpha _{a}}$ and $%
V_{\Psi ^{\ast }\beta ^{a}}$ are%
\begin{equation*}
\left\{ 
\begin{array}{l}
V_{\Psi ^{\ast }\alpha _{a}}\left( g,\eta \right) =\left( 0,-\mathrm{Ad}%
_{g_{-}^{-1}}^{G}\Pi _{\mathfrak{g}_{+}}\mathrm{Ad}_{g_{-}}^{G}T_{a}\right)
_{\left( g,\eta \right) } \\ 
\\ 
V_{\Psi ^{\ast }\beta ^{a}}\left( g,\eta \right) =\left( gT^{a},\mathrm{ad}%
_{T^{a}}^{\mathfrak{g}\ast }\eta -\lambda \mathrm{Ad}_{g}^{G\ast }\hat{c}%
\left( \mathrm{Ad}_{g}^{G}T^{a}\right) \right)%
\end{array}%
\right.
\end{equation*}%
which allows to we calculate the Dirac matrix%
\begin{equation*}
C\left( g,\eta \right) =\left( 
\begin{array}{cc}
C_{\Psi ^{\ast }\alpha _{a},\Psi ^{\ast }\alpha _{b}}\left( g,\eta \right) & 
C_{\Psi ^{\ast }\alpha _{a},\Psi ^{\ast }\beta ^{b}}\left( g,\eta \right) \\ 
-C_{\Psi ^{\ast }\alpha _{a},\Psi ^{\ast }\beta ^{b}}\left( g,\eta \right) & 
C_{\Psi ^{\ast }\beta ^{a},\Psi ^{\ast }\beta ^{b}}\left( g,\eta \right)%
\end{array}%
\right)
\end{equation*}%
determined by the entries 
\begin{eqnarray*}
C_{\Psi ^{\ast }\alpha _{a},\Psi ^{\ast }\alpha _{b}}\left( g,\eta \right)
&=&\left\langle \Psi ^{\ast }\alpha _{a},V_{\Psi ^{\ast }\alpha
_{b}}\right\rangle _{\left( g,\eta \right) }=\left\langle \Psi ^{\ast
}\alpha _{a},\left( 0,w_{\alpha _{b}}\right) \right\rangle _{\left( g,\eta
\right) }=0 \\
&& \\
C_{\Psi ^{\ast }\alpha _{a},\Psi ^{\ast }\beta ^{b}}\left( g,\eta \right)
&=&\left\langle \Psi ^{\ast }\alpha _{a},V_{\Psi ^{\ast }\beta
^{b}}\right\rangle _{\left( g,\eta \right) }=\delta _{a}^{b} \\
&& \\
C_{\Psi ^{\ast }\beta ^{a},\Psi ^{\ast }\beta ^{b}}\left( g,\eta \right)
&=&\left\langle \Psi ^{\ast }\beta ^{a},V_{\Psi ^{\ast }\beta
^{b}}\right\rangle _{\left( g,\eta \right) }=\Omega _{c}^{ab}\left( g,\eta
\right)
\end{eqnarray*}%
Here we wrote 
\begin{equation*}
\Omega _{c}^{ab}\left( g,\eta \right) :=-\left\langle C\left( g^{-1}\right)
+\eta ,\left[ T^{a},T^{b}\right] \right\rangle -c\left( T^{a},T^{b}\right)
\end{equation*}%
The Dirac matrix is then%
\begin{equation*}
C\left( g,\eta \right) =\left( 
\begin{array}{cc}
0_{n\times n} & I_{n\times n} \\ 
-I_{n\times n} & \Omega _{c}\left( g,\eta \right)%
\end{array}%
\right)
\end{equation*}

Now, we are ready to built up the Dirac brackets: for any couple of function 
$\mathcal{F},\mathcal{G}\in C^{\infty }\left( G\times \mathfrak{g}^{\ast
}\right) $, the Dirac bracket gives the restriction of the Poisson bracket
on $G\times \mathfrak{g}^{\ast }$ to the constrained submanifold $\mathcal{N}%
\left( g_{-},\eta _{-}\right) $, and it is defined as 
\begin{equation*}
\begin{array}{l}
\left\{ \mathcal{F},\mathcal{G}\right\} _{c}^{D}\left( g,\eta \right)
=\left\{ \mathcal{F},\mathcal{G}\right\} _{c}\left( g,\eta \right) \\ 
\qquad \qquad \qquad \qquad +\left\langle \eta +C\left( g^{-1}\right) ,\left[
\Pi _{\mathfrak{g}_{-}}\left\{ \mathcal{F},\alpha \right\} ,\Pi _{\mathfrak{g%
}_{-}}\left\{ \alpha ,\mathcal{G}\right\} \right] \right\rangle _{\left(
g,\eta \right) } \\ 
\qquad \qquad \qquad \qquad +\left\langle \left\{ \mathcal{F},\alpha
\right\} ,\left\{ \beta ,\mathcal{G}\right\} \right\rangle -\left\langle
\left\{ \mathcal{F},\beta \right\} ,\left\{ \alpha ,\mathcal{G}\right\}
\right\rangle _{\left( g,\eta \right) } \\ 
\qquad \qquad \qquad \qquad +c\left( \Pi _{\mathfrak{g}_{-}}\left\{ \mathcal{%
F},\alpha \right\} _{c}\left( g,\eta \right) ,\Pi _{\mathfrak{g}_{-}}\left\{
\alpha ,\mathcal{G}\right\} _{c}\left( g,\eta \right) \right)%
\end{array}%
\end{equation*}%
where 
\begin{eqnarray*}
\Pi _{\mathfrak{g}_{-}}\left\{ \mathcal{F},\alpha \right\} &=&\left\{ 
\mathcal{F},\alpha _{a}\right\} \left( g,\eta \right) T^{a}=-\mathrm{Ad}%
_{g_{-}^{-1}}^{G}\Pi _{\mathfrak{g}_{-}}\mathrm{Ad}_{g_{-}}^{G}\delta 
\mathcal{F} \\
&& \\
\Pi _{\mathfrak{g}_{+}}\left\{ \mathcal{F},\beta \right\} \left( g,\eta
\right) &=&\Pi _{\mathfrak{g}_{+}}\left( g\mathbf{d}\mathcal{F}-\mathrm{ad}%
_{\delta \mathcal{F}}^{\mathfrak{g}\ast }\left( C\left( g^{-1}\right) +\eta
\right) +c\left( T^{a},\delta \mathcal{F}\right) T_{a}\right)
\end{eqnarray*}%
Thus, we have the following result.

\begin{description}
\item[Proposition:] \textit{The Dirac bracket on the submanifolds }$\mathcal{%
N}_{c}\left( g_{-},\eta _{-}\right) $ \textit{for any couple of function} $%
\mathcal{F},\mathcal{G}\in C^{\infty }\left( G\times \mathfrak{g}^{\ast
}\right) $ \textit{is\ }%
\begin{equation}
\begin{array}{lll}
\left\{ \mathcal{F},\mathcal{G}\right\} _{c}^{D}\left( g,\eta \right) & = & 
\left\langle g\mathbf{d}\mathcal{F},\mathrm{Ad}_{g_{-}^{-1}}^{G}\Pi _{%
\mathfrak{g}_{+}}\mathrm{Ad}_{g_{-}}^{G}\delta \mathcal{G}\right\rangle
-\left\langle g\mathbf{d}\mathcal{G},\mathrm{Ad}_{g_{-}^{-1}}^{G}\Pi _{%
\mathfrak{g}_{+}}\mathrm{Ad}_{g_{-}}^{G}\delta \mathcal{F}\right\rangle \\ 
&  & -\left\langle \eta ,[\mathrm{Ad}_{g_{-}^{-1}}^{G}\Pi _{\mathfrak{g}_{+}}%
\mathrm{Ad}_{g_{-}}^{G}\delta \mathcal{F},\mathrm{Ad}_{g_{-}^{-1}}^{G}\Pi _{%
\mathfrak{g}_{+}}\mathrm{Ad}_{g_{-}}^{G}\delta \mathcal{G}]\right\rangle \\ 
&  & -\left\langle C\left( g_{+}^{-1}\right) ,[\Pi _{\mathfrak{g}_{+}}%
\mathrm{Ad}_{g_{-}}^{G}\delta \mathcal{F},\Pi _{\mathfrak{g}_{+}}\mathrm{Ad}%
_{g_{-}}^{G}\delta \mathcal{G}]\right\rangle \\ 
&  & -c\left( \Pi _{\mathfrak{g}_{+}}\mathrm{Ad}_{g_{-}}^{G}\delta \mathcal{F%
},\Pi _{\mathfrak{g}_{+}}\mathrm{Ad}_{g_{-}}^{G}\delta \mathcal{G}\right)%
\end{array}
\label{PL brack on N}
\end{equation}%
\textit{It is a} \textit{nondegenerate bracket.}
\end{description}

\bigskip

\begin{description}
\item[Remark I:] \textit{In the particular case} $\left( g_{-},\eta
_{-}\right) =\left( e,0\right) $ \textit{we recover the cotangent bundle of }%
$G_{+}$ \textit{endowed with the Poisson structure} 
\begin{eqnarray*}
\left\{ \mathcal{F},\mathcal{G}\right\} _{c}^{D}\left( g_{+},\eta
_{+}\right) &=&\left\langle g\mathbf{d}\mathcal{F},\Pi _{\mathfrak{g}%
_{+}}\delta \mathcal{G}\right\rangle -\left\langle g\mathbf{d}\mathcal{G}%
,\Pi _{\mathfrak{g}_{+}}\delta \mathcal{F}\right\rangle \\
&&-\left\langle \eta +C\left( g_{+}^{-1}\right) ,[\Pi _{\mathfrak{g}%
_{+}}\delta \mathcal{F},\Pi _{\mathfrak{g}_{+}}\delta \mathcal{G}%
]\right\rangle \\
&&-c\left( \Pi _{\mathfrak{g}_{+}}\delta \mathcal{F},\Pi _{\mathfrak{g}%
_{+}}\delta \mathcal{G}\right)
\end{eqnarray*}
\end{description}

\bigskip

\begin{description}
\item[Remark II:] \textit{Let us suppose that the restriction of the cocycle
on }$G$\textit{\ to }$G_{\pm }$ \textit{is such that}%
\begin{equation*}
\left. C\right\vert _{G_{\pm }}:G_{\pm }\longrightarrow \mathfrak{g}_{\mp
}^{\ast }
\end{equation*}%
\textit{which in turn implies that }%
\begin{equation*}
\left. \hat{c}\right\vert _{\mathfrak{g}_{\pm }}:\mathfrak{g}_{\pm
}\longrightarrow \mathfrak{g}_{\mp }^{\ast }
\end{equation*}%
\textit{then,} \textit{for any couple of function} $\mathcal{F},\mathcal{G}%
\in C^{\infty }\left( G\times \mathfrak{g}^{\ast }\right) $\textit{, the
Dirac bracket on }$\mathcal{N}_{c}\left( g_{-},\eta _{-}\right) $ \textit{of
eq. }$\left( \ref{PL brack on N}\right) $\textit{\ reduces to\ }%
\begin{eqnarray}
\left\{ \mathcal{F},\mathcal{G}\right\} _{c}^{D}\left( g,\eta \right)
&=&\left\langle g\mathbf{d}\mathcal{F},\mathrm{Ad}_{g_{-}^{-1}}^{G}\Pi _{%
\mathfrak{g}_{+}}\mathrm{Ad}_{g_{-}}^{G}\delta \mathcal{G}\right\rangle
-\left\langle g\mathbf{d}\mathcal{G},\mathrm{Ad}_{g_{-}^{-1}}^{G}\Pi _{%
\mathfrak{g}_{+}}\mathrm{Ad}_{g_{-}}^{G}\delta \mathcal{F}\right\rangle
\label{PL brack on N c=0} \\
&&-\left\langle \eta ,[\mathrm{Ad}_{g_{-}^{-1}}^{G}\Pi _{\mathfrak{g}_{+}}%
\mathrm{Ad}_{g_{-}}^{G}\delta \mathcal{F},\mathrm{Ad}_{g_{-}^{-1}}^{G}\Pi _{%
\mathfrak{g}_{+}}Ad_{g_{-}}^{G}\delta \mathcal{G}]\right\rangle  \notag
\end{eqnarray}%
\textit{where there are no traces of the cocycle. Moreover, at }$\left(
g_{-},\eta _{-}\right) =\left( e,0\right) $ \textit{it reduces} $to$%
\begin{eqnarray*}
\left\{ \mathcal{F},\mathcal{G}\right\} _{c}^{D}\left( g_{+},\eta
_{-}\right) &=&\left\langle g\mathbf{d}\mathcal{F},\Pi _{\mathfrak{g}%
_{+}}\delta \mathcal{G}\right\rangle -\left\langle g\mathbf{d}\mathcal{G}%
,\Pi _{\mathfrak{g}_{+}}\delta \mathcal{F}\right\rangle \\
&&-\left\langle \eta ,[\Pi _{\mathfrak{g}_{+}}\delta \mathcal{F},\Pi _{%
\mathfrak{g}_{+}}\delta \mathcal{G}]\right\rangle
\end{eqnarray*}%
\textit{the restriction to the tangent bundle of} $G_{+}$ \textit{equipped
with the canonical symplectic form. The hamiltonian vector field in this
case is }%
\begin{equation}
V_{\mathcal{H}}\left( g,\eta \right) =\left( g\left( \mathrm{Ad}%
_{g_{-}^{-1}}^{G}\Pi _{\mathfrak{g}_{+}}\mathrm{Ad}_{g_{-}}^{G}\delta 
\mathcal{H}\right) ,\mathrm{Ad}_{g_{-}}^{G\ast }\Pi _{\mathfrak{g}_{+}^{\ast
}}\mathrm{Ad}_{g_{-}^{-1}}^{G\ast }\left( \mathrm{ad}_{\mathrm{Ad}%
_{g_{-}^{-1}}^{G}\Pi _{\mathfrak{g}_{+}}\mathrm{Ad}_{g_{-}}^{G}\delta 
\mathcal{H}}^{\mathfrak{g}\ast }\eta \right) \right)
\label{Ham vec dirac brack on N c=0}
\end{equation}
\end{description}

\section{Loop groups}

Loop groups constitutes the main application of the above construction, also
considering their central extension. Thus, let $G=LH$ denotes the set of
maps from $S^{1}$ to the Lie group $H$, and $\mathfrak{g}=L\mathfrak{h}$ the
maps of $S^{1}$ into the Lie algebra $\mathfrak{h}$ of $H$. We assume the $%
\mathfrak{h}$ is equipped with a nondegenerate $\mathrm{Ad}^{H}$-invariant
symmetric bilinear form $\left( ,\right) _{\mathfrak{h}}$.

For $g\in G$, $g^{\prime }$ denotes the derivative in the loop parameter $%
s\in S^{1}$, and we write $vg^{-1}$ and $g^{-1}v$ for the right and left
translation of any vector field $v\in TG$. Frequently we will work with the
dense subset $L\mathfrak{h}^{\ast }\subset $ $(L\mathfrak{h})^{\ast }$%
\textit{\ }instead of $(L\mathfrak{h})^{\ast }$ itself, and we identify it
with $L\mathfrak{h}$ through the map $\psi :L\mathfrak{h}\rightarrow L%
\mathfrak{h}^{\ast }$ provided by the bilinear form 
\begin{equation*}
\left( ,\right) _{\mathfrak{g}}\equiv \dfrac{1}{2\pi }\int_{S^{1}}\left(
,\right) _{\mathfrak{h}}
\end{equation*}%
on $\mathfrak{g}$. In this framework, the two cocycle $c_{\mathrm{k}}:%
\mathfrak{g}\times \mathfrak{g}\longrightarrow \mathbb{R}$ is given by the
bilinear form $\Gamma _{\mathrm{k}}:\mathfrak{g}\times \mathfrak{g}%
\rightarrow \mathbb{R}$ \cite{Pres-Seg}, 
\begin{equation*}
c_{\mathrm{k}}(X,Y)\equiv \Gamma _{\mathrm{k}}(X,Y)=\frac{\mathrm{k}}{2\pi }%
\int_{S^{1}}\left( X\left( s\right) ,Y^{\prime }\left( s\right) \right) _{%
\mathfrak{h}}\,ds
\end{equation*}%
with $X\left( s\right) ,Y\left( s\right) \in \mathfrak{h}$.\ It is derived
from the one cocycle $C_{\mathrm{k}}:G\rightarrow \mathfrak{g}^{\ast }$, 
\begin{equation*}
C_{\mathrm{k}}\left( l\right) =\mathrm{k}\psi \left( l^{\prime }l^{-1}\right)
\end{equation*}%
Observe that is an coadjoint cocycle 
\begin{equation*}
C_{\mathrm{k}}\left( kl\right) =\mathrm{Ad}_{k^{-1}}^{G\ast }C_{\mathrm{k}%
}\left( l\right) +C_{\mathrm{k}}\left( k\right)
\end{equation*}

As above, we assume that $H=H_{+}H_{-}$ with $H_{+},H_{-}$ being Lie
subgroups of $H$, and $\mathfrak{h}=\mathfrak{h}_{+}\oplus \mathfrak{h}_{-}$%
, where $\mathfrak{h}_{+},\mathfrak{h}_{-}$ are Lie subalgebras of\textit{\ }%
$\mathfrak{h}$.\ Moreover, we assume that\textit{\ t}he subspaces $\mathfrak{%
h}_{+},\mathfrak{h}_{-}$ are isotropic in relation with the bilinear form%
\textit{\ }$\left( ,\right) _{\mathfrak{h}}$. Then, the restriction of the
bijection $\psi :\mathfrak{h}\longrightarrow \mathfrak{h}^{\ast }$ to $%
\mathfrak{h}_{\pm }$ provides the identification $\psi :\mathfrak{h}_{\pm
}\longrightarrow \mathfrak{h}_{\mp }^{\ast }$, and the restriction of the
cocycle to the factors $G_{\pm }$ is then the map $C_{\mathrm{k}}:G_{\pm
}\rightarrow \mathfrak{h}_{\mp }^{\ast }$ . Moreover, the bilinear form $%
\left( ,\right) _{\mathfrak{h}}$ and the 2 cocycle\textit{\ }$c_{\mathrm{k}}:%
\mathfrak{g}\times \mathfrak{g}\longrightarrow \mathbb{R}$ restricted to $%
\mathfrak{g}_{\pm }$ vanish, falling in the situation of \emph{Remark II} of
the previous section.

Thus, the Dirac bracket on the fiber $\mathcal{N}\left( g_{-},\eta
_{-}\right) $ coincides with $\left( \ref{PL brack on N c=0}\right) $. This
is the framework for the developments in the second part of this work.

\section{The left action of $G$ on $G\times \mathfrak{g}^{\mathfrak{\ast }}$
and its restriction to $\mathcal{N}\left( g_{-},\protect\eta _{-}\right) $}

This section is devoted to study the left action of $G$ on $G\times 
\mathfrak{g}^{\mathfrak{\ast }}$and how it restricts to the fibers $\mathcal{%
N}\left( g_{-},\eta _{-}\right) $, analyzing the momentum maps borrowed from
the phase space $G\times \mathfrak{g}^{\mathfrak{\ast }}$ supplied with the
canonical symplectic structure.

The selected cocycle breaks explicitly the left action symmetry then it is
no longer an endomorphism on $\left( T^{\ast }G,\omega _{c}\right) $.
However, since the cocycle contribution seems to disappear from the Dirac
brackets, we wonder if that symmetry would be restored on some the fibers $%
\mathcal{N}\left( g_{-},\eta _{-}\right) $. If it were the case, the
infinitesimal generator $X_{T^{\ast }G}$ would be related to the momentum
functions associated with the canonical symplectic form, so we consider the
momentum map associated to left translation on $\left( T^{\ast }G,\omega _{%
\mathtt{o}}\right) $, namely $J_{B}^{L}:T^{\ast }G\longrightarrow \mathfrak{g%
}^{\ast }$ defined as 
\begin{equation*}
J_{B}^{L}(g,\eta )=\mathrm{Ad}_{g^{-1}}^{G\ast }\eta
\end{equation*}%
It is worth to stress that it is not a momentum map for $\left( T^{\ast
}G,\omega _{c}\right) $. Despite of this fact, we shall consider the
associated momentum function $j_{X}^{L}$, namely%
\begin{equation*}
j_{X}^{L}\left( g,\eta \right) =\left\langle \mathrm{Ad}_{g^{-1}}^{G\ast
}\eta ,X\right\rangle =\left\langle \eta ,\mathrm{Ad}_{g^{-1}}^{G}X\right%
\rangle
\end{equation*}%
satisfying $\imath _{X_{T^{\ast }G}}\omega _{\mathtt{o}}=dj_{X}^{L}$, and
construct the associated hamiltonian vector fields. In doing so, we need the
differential of $j_{X}^{L}$,%
\begin{equation*}
dj_{X}^{L}=\left( g^{-1}\mathrm{ad}_{Ad_{g^{-1}}X}^{\mathfrak{g}\ast }\eta ,%
\mathrm{Ad}_{g^{-1}}^{G}X\right)
\end{equation*}%
So, for an arbitrary function $\mathcal{F}$ on $G\times \mathfrak{g}^{\ast }$%
, its Lie derivative along the hamiltonian vector field of $j_{X}^{L}$
projected on the tangent space to $\mathcal{N}\left( g_{-},\eta _{-}\right) $
is given by the Dirac bracket

\begin{eqnarray*}
\left\{ \mathcal{F},j_{X}^{L}\right\} _{c}^{D}\left( g,\eta \right)
&=&\left\langle g\mathbf{d}\mathcal{F},\mathrm{Ad}_{g_{-}^{-1}}^{G}\Pi _{%
\mathfrak{g}_{+}}\mathrm{Ad}_{g_{+}^{-1}}^{G}X\right\rangle \\
&&-\left\langle \mathrm{ad}_{Ad_{g^{-1}}X}^{\mathfrak{g}\ast }\eta ,\mathrm{%
Ad}_{g_{-}^{-1}}^{G}\Pi _{\mathfrak{g}_{+}}\mathrm{Ad}_{g_{-}}^{G}\delta 
\mathcal{F}\right\rangle \\
&&-\left\langle \eta ,[\mathrm{Ad}_{g_{-}^{-1}}^{G}\Pi _{\mathfrak{g}_{+}}%
\mathrm{Ad}_{g_{-}}^{G}\delta \mathcal{F},\mathrm{Ad}_{g_{-}^{-1}}^{G}\Pi _{%
\mathfrak{g}_{+}}\mathrm{Ad}_{g_{+}^{-1}}^{G}X]\right\rangle
\end{eqnarray*}%
so, the hamiltonian vector field is 
\begin{equation*}
V_{j_{X}^{_{L}}}\left( g,\eta \right) =\left( g\mathrm{Ad}%
_{g_{-}^{-1}}^{G}\Pi _{\mathfrak{g}_{+}}\mathrm{Ad}_{g_{+}^{-1}}^{G}X,-%
\mathrm{Ad}_{g_{-}}^{G\ast }\Pi _{\mathfrak{g}_{+}^{\ast }}\mathrm{Ad}%
_{g_{-}^{-1}}^{G\ast }ad_{\mathrm{Ad}_{g_{-}^{-1}}^{G}\Pi _{\mathfrak{g}_{-}}%
\mathrm{Ad}_{g_{+}^{-1}}^{G}X}^{\mathfrak{g}\ast }\eta \right)
\end{equation*}%
For the fiber on $\left( e,0\right) $ and $X=X_{+}\in \mathfrak{g}_{+}$ we
get%
\begin{equation*}
V_{j_{X_{+}}^{_{L}}}\left( g,\eta \right) =\left( X_{+}g_{+},0\right)
\end{equation*}%
that is just the infinitesimal generator of the left action of $G_{+}$ on $%
T^{\ast }G_{+}$.

A test to see if the symmetry is restored on $\mathcal{N}\left( g_{-},\eta
_{-}\right) $ is to calculate the Dirac bracket of two momentum functions.
If the result is that the momentum function close a Lie algebra under the
Dirac bracket, then we have a Lie algebra morphism between the Lie algebra
of the group and the Lie algebra of momentum functions, showing that the
symmetry is symplectically realized on $\mathcal{N}\left( g_{-},\eta
_{-}\right) $.

\begin{description}
\item[Proposition:] \textit{Let }$\eta _{-}$ \textit{a} \emph{character} 
\textit{of} $\mathfrak{g}_{-}$. \textit{Then, the map }$\mathfrak{g}%
\longrightarrow C^{\infty }\left( \mathcal{N}\left( g_{-},\eta _{-}\right)
\right) $ \textit{such that }$X\longrightarrow $\textit{\ }$j_{X}^{L}$ 
\textit{is a Lie algebra homomorphism in relation with the Dirac bracket }$%
\left( \ref{PL brack on N c=0}\right) $.\textit{in }$\mathcal{N}\left(
g_{-},\eta _{-}\right) $.
\end{description}

\textbf{Proof: }The Dirac bracket between momentum functions is%
\begin{eqnarray*}
\left\{ j_{X}^{L},j_{Y}^{L}\right\} _{c}^{D}\left( g,\eta \right)
&=&\left\langle \eta ,\left[ \mathrm{Ad}_{g^{-1}}^{G}X,\mathrm{Ad}%
_{g_{-}^{-1}}^{G}\Pi _{\mathfrak{g}_{+}}\mathrm{Ad}_{g_{+}^{-1}}^{G}Y\right]
\right\rangle \\
&&+\left\langle \eta ,\left[ \mathrm{Ad}_{g_{-}^{-1}}^{G}\Pi _{\mathfrak{g}%
_{+}}\mathrm{Ad}_{g_{+}^{-1}}^{G}X,\mathrm{Ad}_{g_{-}^{-1}}^{G}\Pi _{%
\mathfrak{g}_{-}}\mathrm{Ad}_{g_{+}^{-1}}^{G}Y\right] \right\rangle
\end{eqnarray*}%
Observe that if $\eta _{-}$ is a \emph{character} of $\mathfrak{g}_{-}$ we
can add a term 
\begin{equation*}
0=\left\langle \eta ,\left[ \mathrm{Ad}_{g_{-}^{-1}}^{G}\Pi _{\mathfrak{g}%
_{-}}\mathrm{Ad}_{g_{+}^{-1}}^{G}X,\mathrm{Ad}_{g_{-}^{-1}}^{G}\Pi _{%
\mathfrak{g}_{-}}\mathrm{Ad}_{g_{+}^{-1}}^{G}Y\right] \right\rangle
\end{equation*}%
so $\left\{ j_{X}^{L},j_{Y}^{L}\right\} _{c}^{D}$ turns in%
\begin{equation*}
\left\{ j_{X}^{L},j_{Y}^{L}\right\} _{c}^{D}\left( g,\eta \right)
=\left\langle \eta ,\left[ \mathrm{Ad}_{g^{-1}}^{G}X,\mathrm{Ad}%
_{g^{-1}}^{G}Y\right] \right\rangle
\end{equation*}%
that is equivalent to%
\begin{equation*}
\left\{ j_{X_{+}}^{L},j_{Y}^{L}\right\} _{c}^{D}\left( g,\eta \right) =j_{ 
\left[ X,Y\right] }^{L}\left( g,\eta \right)
\end{equation*}%
indicating that the left invariance under the action of $G_{+}$ is restored
on $\mathcal{N}\left( g_{-},\eta _{-}\right) $ when $\eta _{-}$ is a \emph{%
character} of $\mathfrak{g}_{-}$.$\blacksquare $

\subsection{The action of the centrally extended group $G^{\wedge }$}

Now, we consider the action of the centrally extended loop group $G^{\wedge
} $ which is more frequently involved in the hamiltonian framework of loop
groups with extended symplectic form concerning WZNW models, rather than the
simple action of the group itself. So, it seems natural to work out the
symmetries generated in this case by repeating the developments of the
previous section.

Let $\mathfrak{g}_{c}$ the \emph{centrally extended} Lie algebra $\mathfrak{g%
}$ defined by the cocycle $c:\mathfrak{g}\otimes \mathfrak{g}\rightarrow 
\mathbb{R}$, and $\mathfrak{g}_{c}^{\ast }$ its dual algebra. The adjoint
and coadjoint actions of $G$ coincides with those of the \emph{centrally
extended} Lie group $G^{\wedge }$ since the action of the extension factor
on $\mathfrak{g}_{c}$ is trivial. Then, because $\mathbb{R}$ acts trivially,
the adjoint action are%
\begin{equation*}
\mathrm{Ad}_{\left( g,b\right) }^{G\wedge }\left( X,a\right) :=\mathrm{Ad}%
_{g}^{G}\left( X,a\right)
\end{equation*}%
Explicitly we have the formulas 
\begin{equation*}
\left\{ 
\begin{array}{l}
\mathrm{Ad}_{g}^{G}\left( X,a\right) :=\left( \mathrm{Ad}_{g}^{G}X,a+\left%
\langle C\left( g^{-1}\right) ,X\right\rangle \right) \\ 
\mathrm{Ad}_{g^{-1}}^{G\ast }\left( \xi ,b\right) :=\left( \mathrm{Ad}%
_{g^{-1}}^{G\ast }\xi +bC\left( g\right) ,b\right) \\ 
\mathrm{ad}_{X}^{\mathfrak{g}}\left( Y,a\right) :=\left( \left[ X,Y\right]
,c\left( X,Y\right) \right)%
\end{array}%
\right.
\end{equation*}

Let us now introduce the action by left translations of $G^{\wedge }$ on $%
G\times \mathfrak{g}^{\ast }$ by defining the \emph{centrally extended
momentum map} 
\begin{equation}
J_{B}^{L\wedge }(g,\eta ):=\mathrm{Ad}_{g^{-1}}^{G\ast \wedge }\left( \eta
,1\right) =\left( \mathrm{Ad}_{g^{-1}}^{G\ast }\eta +C\left( g\right)
,\,1\right)  \label{L5}
\end{equation}%
The momentum function $j_{X}^{L\wedge }$ associated with $J_{B}^{L\wedge }$,
namely%
\begin{equation*}
j_{\left( X,a\right) }^{L\wedge }\left( g,\eta \right) =\left\langle \mathrm{%
Ad}_{g^{-1}}^{G\ast \wedge }\eta ,\left( X,a\right) \right\rangle
=j_{X}^{L}\left( g,\eta \right) +\left\langle C\left( g\right)
,X\right\rangle +a
\end{equation*}%
then 
\begin{equation*}
dj_{\left( X,a\right) }^{L\wedge }=\left( g^{-1}\left( \mathrm{ad}%
_{Ad_{g^{-1}}X}^{\mathfrak{g}\ast }\eta +\hat{c}\left( \mathrm{Ad}%
_{g^{-1}}^{G}X\right) \right) ,\mathrm{Ad}_{g^{-1}}^{G}X\right)
\end{equation*}

The centrally extended canonical symplectic form $\left( \ref{cent ext symp
form}\right) $ gives rise to the hamiltonian vector field 
\begin{equation*}
V_{j_{\left( X,b\right) }^{L\wedge }}\left( g,\eta \right) =\left( Xg,%
\mathrm{ad}_{Ad_{g^{-1}}^{G}X}^{\mathfrak{g}\ast }C\left( g^{-1}\right)
\right)
\end{equation*}%
and Poisson bracket between momentum functions is 
\begin{equation*}
\left\{ j_{\left( X,a\right) }^{L\wedge },j_{\left( Y,b\right) }^{L\wedge
}\right\} _{c}\left( g,\eta \right) =\left\langle dj_{\left( X,a\right)
}^{L\wedge },V_{j_{\left( Y,b\right) }^{L\wedge }}\right\rangle _{\left(
g,\eta \right) }
\end{equation*}%
and the explicit calculation gives%
\begin{equation*}
\left\{ j_{\left( X,a\right) }^{L\wedge },j_{\left( Y,b\right) }^{L\wedge
}\right\} _{c}\left( g,\eta \right) =j_{\left[ \left( X,a\right) ,\left(
Y,b\right) \right] }^{L\wedge }\left( g,\eta \right) +\left\langle C\left(
g\right) ,\left[ X,Y\right] \right\rangle
\end{equation*}%
which reflects the non invariance of the symplectic form $\omega _{c}$.

Now, let us study the behavior of the hamiltonian vector fields $%
V_{j_{\left( X,a\right) }^{L\wedge }}\equiv V_{\left( X,a\right) }^{\mathcal{%
N}}$ on the fibers $\mathcal{N}\left( g_{-},\eta _{-}\right) $ by
calculating the Dirac brackets of these momentum functions. To start with,
let us consider the Dirac bracket with an arbitrary function $\mathcal{F}$, 
\begin{eqnarray*}
\left\{ \mathcal{F},j_{\left( X,a\right) }^{L\wedge }\right\} _{c}^{D}\left(
g,\eta \right) &=&\left\langle g\mathbf{d}\mathcal{F},\mathrm{Ad}%
_{g_{-}^{-1}}^{G}\Pi _{\mathfrak{g}_{+}}\mathrm{Ad}_{g_{+}^{-1}}^{G}X\right%
\rangle \\
&&-\left\langle \mathrm{ad}_{\mathrm{Ad}_{g^{-1}}^{G}X}^{\mathfrak{g}\ast
}\eta +\hat{c}\left( \mathrm{Ad}_{g^{-1}}^{G}X\right) ,\mathrm{Ad}%
_{g_{-}^{-1}}^{G}\Pi _{\mathfrak{g}_{+}}\mathrm{Ad}_{g_{-}}^{G}\delta 
\mathcal{F}\right\rangle \\
&&-\left\langle \eta ,[\mathrm{Ad}_{g_{-}^{-1}}^{G}\Pi _{\mathfrak{g}_{+}}%
\mathrm{Ad}_{g_{-}}^{G}\delta \mathcal{F},\mathrm{Ad}_{g_{-}^{-1}}^{G}\Pi _{%
\mathfrak{g}_{+}}\mathrm{Ad}_{g_{+}^{-1}}^{G}X]\right\rangle
\end{eqnarray*}%
from where we get the hamiltonian vector field%
\begin{eqnarray}
V_{\left( X,a\right) }^{\mathcal{N}}\left( g,\eta \right) &=&\left( g\mathrm{%
Ad}_{g_{-}^{-1}}^{G}\Pi _{\mathfrak{g}_{+}}\mathrm{Ad}_{g_{+}^{-1}}^{G}X,%
\right.  \label{L7} \\
&&\left. -\mathrm{Ad}_{g_{-}}^{G\ast }\Pi _{\mathfrak{g}_{+}^{\ast }}\mathrm{%
Ad}_{g_{-}^{-1}}^{G\ast }\left( \mathrm{ad}_{\mathrm{Ad}_{g_{-}^{-1}}^{G}\Pi
_{\mathfrak{g}_{-}}\mathrm{Ad}_{g_{+}^{-1}}^{G}X}^{\mathfrak{g}\ast }\eta +%
\hat{c}\left( \mathrm{Ad}_{g^{-1}}^{G}X\right) \right) \right)  \notag
\end{eqnarray}

In analogous way as in the proposition at the end of the previous section,
it is easy to see the following result.

\begin{description}
\item[Proposition:] \textit{Let }$\left( g_{-},\eta _{-}\right) \in \ker
C\times \mathrm{Char}\left( \mathfrak{g}_{-}\right) $\textit{, then the map }%
$\mathfrak{g}^{\wedge }\longrightarrow C^{\infty }\left( \mathcal{N}\left(
g_{-},\eta _{-}\right) \right) $ \textit{such that }$\left( X,a\right)
\longrightarrow $\textit{\ }$j_{\left( X,a\right) }^{_{L\wedge }}$ \textit{%
is a Lie algebra homomorphism in relation with the Dirac bracket }$\left( %
\ref{PL brack on N c=0}\right) $ \textit{on }$C^{\infty }\left( \mathcal{N}%
\left( g_{-},\eta _{-}\right) \right) $\textit{, namely}%
\begin{equation*}
\left\{ j_{\left( X,a\right) }^{L\wedge },j_{\left( Y,b\right) }^{L\wedge
}\right\} _{c}^{D}\left( g,\eta \right) =j_{\left[ \left( X,a\right) ,\left(
Y,b\right) \right] }^{L\wedge }\left( g,\eta \right)
\end{equation*}
\end{description}

In terms of hamiltonian vector fields it means that 
\begin{equation*}
\left[ V_{\left( X,a\right) }^{\mathcal{N}},V_{\left( Y,b\right) }^{\mathcal{%
N}}\right] =-V_{\left[ \left( X,a\right) ,\left( Y,b\right) \right] }^{%
\mathcal{N}}
\end{equation*}%
so, the linear map $\mathfrak{g}^{\wedge }\longrightarrow \mathfrak{X}\left( 
\mathcal{N}\left( g_{-},\eta _{-}\right) \right) $ / $\left( X,a\right)
\longrightarrow V_{\left( X,a\right) }^{\mathcal{N}}$ is a Lie algebra
antihomomorphism showing that it is the infinitesimal generator of a well
defined symplectic left action of $G^{\wedge }$ on $\mathcal{N}\left(
g_{-},\eta _{-}\right) $.

In analogous way as in ref. \cite{CapMon JMP} in the framework of $\left(
T^{\ast }G,\omega _{\mathtt{o}}\right) $, this infinitesimal action of $%
\mathfrak{g}$ on\textit{\ }$\mathcal{N}\left( g_{-},\eta _{-}\right) $ gives
rise to a finite action of $G^{\wedge }$ on $\mathcal{N}\left( g_{-},\eta
_{-}\right) $.

\begin{description}
\item[Proposition:] \textit{The vector field }$V_{\left( X,a\right) }^{%
\mathcal{N}}\in \mathfrak{X}\left( \mathcal{N}\left( g_{-},\eta _{-}\right)
\right) $\textit{,} \textit{for }$X\in \mathfrak{g}$\textit{\ and }$\eta
_{-} $ \textit{a character of }$\mathfrak{g}_{-}$\textit{, is the
infinitesimal generator associated with the action }$\mathsf{d:}G^{\wedge
}\times \mathcal{N}\left( g_{-},\eta _{-}\right) \longrightarrow \mathcal{N}%
\left( g_{-},\eta _{-}\right) $ \textit{defined as}%
\begin{eqnarray}
&&\mathsf{d}\left( \left( h,a\right) ,\left( g,\eta \right) \right)
\label{SymplecticInducedAction} \\
&=&\left( g\mathrm{Ad}_{g_{-}^{-1}}^{G}\Pi _{\mathfrak{g}_{+}}\left(
g_{+}^{-1}hg_{+}\right) \right. ,  \notag \\
&&\left. \mathrm{Ad}_{g_{-}}^{G\ast }\Pi _{\mathfrak{g}_{+}^{\ast }}\left( 
\mathrm{Ad}_{\Pi _{G_{-}}\left( g_{+}^{-1}hg_{+}\right) }^{G\ast }\mathrm{Ad}%
_{g_{-}}^{G\ast }\eta +C\left( \left( \Pi _{G_{-}}\left(
g_{+}^{-1}hg_{+}\right) \right) ^{-1}\right) \right) \right)  \notag
\end{eqnarray}%
$\forall $ $\left( g,\eta \right) =\left( g_{+}g_{-},\eta _{+}+\eta
_{-}\right) \in \mathcal{N}\left( g_{-},\eta _{-}\right) $.
\end{description}

\textbf{Proof: }It follows by straightforward calculation of the
differential of this map.$\blacksquare $

\begin{description}
\item[Remark:] \textit{Observe that for} $g_{-}=e$ \textit{and} $\eta _{-}=0$
\textit{it turns into}%
\begin{eqnarray*}
&&\mathsf{d}\left( \left( h,a\right) ,\left( g,\eta \right) \right) \\
&=&\left( g_{+}\Pi _{\mathfrak{g}_{+}}\left( g_{+}^{-1}hg_{+}\right) \right.
, \\
&&\left. \Pi _{\mathfrak{g}_{+}^{\ast }}\left( \mathrm{Ad}_{\Pi
_{G_{-}}\left( g_{+}^{-1}hg_{+}\right) }^{G\ast }\eta _{+}+C\left( \left(
\Pi _{G_{-}}\left( g_{+}^{-1}hg_{+}\right) \right) ^{-1}\right) \right)
\right)
\end{eqnarray*}%
\textit{This action was introduced in} \cite{CM},\cite{CMZ} \textit{as the
fundamental ingredient underlying the hamiltonian Poisson Lie }$T$\textit{%
-duality scheme.}
\end{description}

In so far we have then show that, despite there are no left translation
symmetry on the phase space $\left( T^{\ast }G,\omega _{c}\right) $, it is
restored on some particular fibers $\mathcal{N}\left( g_{-},\eta _{-}\right) 
$ turning them into potentially interesting phase space for systems
symmetric under the projection on $\mathcal{N}\left( g_{-},\eta _{-}\right) $
of the left action of $G^{\wedge }$ on $T^{\ast }G$.

\section{The Hamilton equations and collective dynamics}

In order to study dynamical systems with the kind of phase space described
above, we explicitly write the Hamilton equations in the whole space $%
T^{\ast }G$ and those on the constrained submanifolds $\mathcal{N}\left(
g_{-},\eta _{-}\right) $ produced by the Dirac brackets, both for a generic
Hamilton function $\mathcal{H}$ on $T^{\ast }G$.

The Hamilton equations on $T^{\ast }G$ are determined by the hamiltonian
vector field associated to a Hamilton function $\mathcal{H}$ through the
symplectic form $\left( \ref{cent ext symp form}\right) $, that was given in 
$\left( \ref{ham vect cent exten sympl form}\right) $. So, defining $d%
\mathcal{H}=\left( \mathbf{d}\mathcal{H},\delta \mathcal{H}\right) \in
T_{\left( g,\eta \right) }^{\ast }\left( G\times \mathfrak{g}^{\ast }\right) 
$, these Hamilton equation are%
\begin{equation*}
\left\{ 
\begin{array}{l}
g^{-1}\dot{g}=\delta \mathcal{H} \\ 
\\ 
\dot{\eta}=\mathrm{ad}_{\delta \mathcal{H}}^{\mathfrak{g}\ast }\eta -g%
\mathbf{d}\mathcal{H}+\mathrm{Ad}_{g}^{G\ast }\hat{c}\left( \mathrm{Ad}%
_{g}^{G}\delta \mathcal{H}\right)%
\end{array}%
\right.
\end{equation*}

From the hamiltonian vector field $\left( \ref{Ham vec dirac brack on N c=0}%
\right) $, which is tangent to $\mathcal{N}\left( g_{-},\eta _{-}\right) $,
we get the Hamilton equations on this phase space: 
\begin{equation}
\left\{ 
\begin{array}{l}
g^{-1}\dot{g}=\mathrm{Ad}_{g_{-}^{-1}}^{G}\Pi _{\mathfrak{g}_{+}}\mathrm{Ad}%
_{g_{-}}^{G}\delta \mathcal{H} \\ 
\\ 
\dot{\eta}=\mathrm{Ad}_{g_{-}}^{G\ast }\Pi _{\mathfrak{g}_{+}^{\ast }}%
\mathrm{Ad}_{g_{-}^{-1}}^{G\ast }\left( \mathrm{ad}_{\mathrm{Ad}%
_{g_{-}^{-1}}^{G}\Pi _{\mathfrak{g}_{+}}\mathrm{Ad}_{g_{-}}^{G}\delta 
\mathcal{H}}^{\mathfrak{g}\ast }\eta -g\mathbf{d}\mathcal{H}\right)%
\end{array}%
\right.  \label{Dirac Ham equations}
\end{equation}%
In terms of the components in $G_{\pm }$ and $\mathfrak{g}_{\pm }^{\ast }$
they means%
\begin{equation*}
\left\{ 
\begin{array}{l}
g_{+}^{-1}\dot{g}_{+}=\Pi _{\mathfrak{g}_{+}}\mathrm{Ad}_{g_{-}}^{G}\delta 
\mathcal{H} \\ 
\\ 
\dot{g}_{-}=0 \\ 
\\ 
\dot{\eta}_{+}=\mathrm{Ad}_{g_{-}}^{G\ast }\Pi _{\mathfrak{g}_{+}^{\ast }}%
\mathrm{Ad}_{g_{-}^{-1}}^{G\ast }\left( \mathrm{ad}_{\mathrm{Ad}%
_{g_{-}^{-1}}^{G}\Pi _{\mathfrak{g}_{+}}\mathrm{Ad}_{g_{-}}^{G}\delta 
\mathcal{H}}^{\mathfrak{g}\ast }\eta -g\mathbf{d}\mathcal{H}\right) \\ 
\\ 
\dot{\eta}_{-}=0%
\end{array}%
\right.
\end{equation*}

We shall be concerned with a dynamical system ruled by a Hamilton function
of the type 
\begin{equation*}
\begin{array}{ccc}
\mathcal{H}=\mathsf{h}\circ J_{B}^{L\wedge } & / & \mathsf{h}:\mathfrak{g}%
^{\wedge \ast }\longrightarrow \mathbb{R}%
\end{array}%
\end{equation*}%
with $J_{B}^{L\wedge }$ being the momentum map associated with the centrally
extended left translation on $\left( T^{\ast }G,\omega _{\mathtt{o}}\right) $
given in eq. $\left( \ref{L5}\right) $. Despite it is apparently of
collective type, actually it is not the case because $J_{B}^{L\wedge }$ is
fails in to be a momentum map for $\left( T^{\ast }G,\omega _{c}\right) $,
which lacks of left invariance symmetry.

However, it is a truly collective system on $\mathcal{N}\left( g_{-},\eta
_{-}\right) $, for $\left( g_{-},\eta _{-}\right) \in \ker C\times \mathrm{%
Char}\left( \mathfrak{g}_{-}\right) $, where the centrally extended left
symmetry is restored. In these fibers, the dynamics acquires interesting
properties which we briefly recall now \cite{Guillemin-Sternberg}. The
derivative of some function $\mathcal{F}$ along the flux of the hamiltonian
vector field $\left( \ref{Ham vec dirac brack on N c=0}\right) $ is%
\begin{equation}
\mathcal{\dot{F}}=\mathbf{L}_{V_{\mathcal{H}}}\mathcal{F}=\left\{ \mathcal{F}%
,\mathcal{H}\right\} _{c}^{D}  \label{F dot}
\end{equation}%
Let us denote by $\varphi _{\left( g,\eta \right) }:G^{\wedge
}\longrightarrow \mathcal{N}\left( g_{-},\eta _{-}\right) $ the orbit map
associated with the action $\left( \ref{SymplecticInducedAction}\right) $
such that 
\begin{equation*}
\varphi _{\left( g,\eta \right) }\left( h,a\right) =\mathsf{d}\left( \left(
h,a\right) ,\left( g,\eta \right) \right)
\end{equation*}%
then, its differential at the neutral element of $G^{\wedge }$ is the map $%
\varphi _{\left( g,\eta \right) \ast }:\mathfrak{g}^{\wedge }\longrightarrow
T_{\left( g,\eta \right) }\mathcal{N}\left( g_{-},\eta _{-}\right) $ gives
rise to the infinitesimal generator $V_{\left( X,a\right) }^{\mathcal{N}}$
on $\mathcal{N}\left( g_{-},\eta _{-}\right) $ defined in $\left( \ref{L7}%
\right) $, then 
\begin{equation*}
\varphi _{\left( g,\eta \right) \ast }\left( X,a\right) :=V_{\left(
X,a\right) }^{\mathcal{N}}\left( g,\eta \right)
\end{equation*}

The Hamiltonian vector field associated to $\mathcal{H}=\mathsf{h}\circ
J_{B}^{L\wedge }$ is then\textit{\ } 
\begin{equation*}
V_{\mathcal{H}}\left( g,\eta \right) =\varphi _{\left( g,\eta \right) \ast }%
\mathcal{L}_{\mathsf{h}}\left( J_{B}^{L\wedge }\left( g,\eta \right) \right)
\end{equation*}%
where $\mathcal{L}_{\mathsf{h}}:\mathfrak{g}^{\wedge \ast }\longrightarrow 
\mathfrak{g}^{\wedge }$ is defined as 
\begin{equation*}
\left\langle \xi ,\mathcal{L}_{\mathsf{h}}\left( \eta ,b\right)
\right\rangle :=\left. \frac{d}{dt}\mathsf{h}\left( \eta +t\xi ,b\right)
\right\vert _{t=0}
\end{equation*}%
meaning that $V_{\mathcal{H}}\left( g,\eta \right) $ coincides with the
infinitesimal generator associated with $\mathcal{L}_{\mathsf{h}}\left(
J_{B}^{L\wedge }\left( g,\eta \right) \right) \in \mathfrak{g}^{\wedge }$.
So, $\left( \ref{F dot}\right) $ is 
\begin{equation*}
\mathcal{\dot{F}}=\left\{ \mathcal{F},j_{\mathcal{L}_{\mathsf{h}}\circ
J_{B}^{L\wedge }}^{L\wedge }\right\} _{c}^{D}
\end{equation*}

In this way, at least locally, the integral curves of the hamiltonian vector
field are orbits $\mathsf{d}\left( \left( h\left( t\right) ,a\right) ,\left(
g,\eta \right) \right) $ of a curve $\gamma :\mathbb{R\longrightarrow }G~/~t%
\overset{\gamma }{\longmapsto }h\left( t\right) $. By the way, this curve is
the same that solves the problem 
\begin{equation*}
\frac{d}{dt}\left( \eta ,b\right) =\mathrm{ad}_{\mathcal{L}_{\mathsf{h}%
}\left( \eta ,b\right) }^{\mathfrak{g}\wedge \ast }\left( \eta ,b\right)
\end{equation*}%
which is equivalent to this one 
\begin{equation*}
\left\{ 
\begin{array}{l}
\left( \eta ,b\right) \left( t\right) =\mathrm{Ad}_{\left( h\left( t\right)
,a\right) }^{G\wedge \ast }\left( \eta ,b\right) \left( t_{\mathtt{o}}\right)
\\ 
\\ 
\left( h\left( t\right) ,a\right) ^{-1}\dfrac{d}{dt}\left( h\left( t\right)
,a\right) =\mathcal{L}_{\mathsf{h}}\left( \eta ,b\right)%
\end{array}%
\right.
\end{equation*}%
Thus, the dynamical system on $\mathcal{N}\left( g_{-},\eta _{-}\right) $ is
replicated on the some coadjoint orbit in $\mathfrak{g}^{\wedge \ast }$, and
the $\mathrm{Ad}$-equivariance for the momentum maps allows to translate
orbits in $\mathfrak{g}^{\wedge \ast }$ into orbits in $\mathcal{N}\left(
g_{-},\eta _{-}\right) $.

\part{From WZNW type model to the Poisson-Lie $\protect\sigma $-model}

As the main application of the results of the first part, we study a
hamiltonian systems which turns into collective type on the phase subspaces $%
\mathcal{N}\left( g_{-},\eta _{-}\right) $, for $\left( g_{-},\eta
_{-}\right) \in \ker C\times \mathrm{Char}\left( \mathfrak{g}_{-}\right) $.
By deriving the Hamilton equations, we retrieve the Lagrange version of the
model, showing that it corresponds to a generalization of the so called 
\emph{Poisson-Lie} $\sigma $\emph{-model.}

\section{A collective Hamilton function on $\mathcal{N}\left( g_{-},\protect%
\eta _{-}\right) $}

Motivated by the previous observation, we now propose to study a hamiltonian
system ruled by the Hamilton function 
\begin{equation}
\mathcal{H}\left( g,\eta \right) =\frac{1}{2}\left( \psi \left(
J_{B}^{L\wedge }\left( g,\eta \right) \right) ,\mathcal{E}\psi \left(
J_{B}^{L\wedge }\left( g,\eta \right) \right) \right) _{\mathfrak{g}^{\wedge
}}  \label{w2}
\end{equation}%
where $J_{B}^{L\wedge }(g,\eta )$ is the centrally extended momentum map
associated with left translations in the phase space $\left( T^{\ast
}G,\omega _{\mathtt{o}}\right) $, introduced in eq. $\left( \ref{L5}\right) $%
. The linear operator $\mathcal{E}:\mathfrak{g}\rightarrow \mathfrak{g}$ is
idempotent $\mathcal{E}^{2}=Id$, and breaks the Ad-invariance of the
bilinear form. For further issues we introduce the operator $\mathcal{E}_{g}:%
\mathfrak{g}\rightarrow \mathfrak{g}$.

\begin{description}
\item[Definition:] \textit{Let the operator }$\mathcal{E}_{g}:\mathfrak{g}%
\rightarrow \mathfrak{g}$ \textit{be defined as}%
\begin{equation*}
\mathcal{E}_{g}=\mathrm{Ad}_{g^{-1}}^{G}\mathcal{E}\mathrm{Ad}%
_{g}^{G}=\left( 
\begin{array}{cc}
-\mathcal{G}_{g}^{-1}\mathcal{B}_{g} & \mathcal{G}_{g}^{-1} \\ 
\mathcal{G}_{g}-\mathcal{B}_{g}\mathcal{G}_{g}^{-1}\mathcal{B}_{g} & 
\mathcal{B}_{g}\mathcal{G}_{g}^{-1}%
\end{array}%
\right)
\end{equation*}%
\textit{where} 
\begin{equation*}
\left\{ 
\begin{array}{l}
\mathcal{G}_{g}=(\Pi _{\mathfrak{g}_{+}}\mathcal{E}_{g}\Pi _{\mathfrak{g}%
_{-}})^{-1}:\mathfrak{g}_{+}\longrightarrow \mathfrak{g}_{-}~/~\left\langle 
\mathcal{G}_{g}X,Y\right\rangle =\left\langle X,\mathcal{G}_{g}Y\right\rangle
\\ 
\\ 
\mathcal{B}_{g}=-\mathcal{G}_{g}\circ \Pi _{\mathfrak{g}_{+}}\mathcal{E}%
_{g}\Pi _{\mathfrak{g}_{+}}:\mathfrak{g}_{+}\longrightarrow \mathfrak{g}%
_{-}~/~\left\langle \mathcal{B}_{g}X,Y\right\rangle =-\left\langle X,%
\mathcal{B}_{g}Y\right\rangle%
\end{array}%
\right.
\end{equation*}%
\textit{which makes} $\mathcal{E}_{g}$ \textit{symmetric} 
\begin{equation*}
\left( \left( Y,\xi \right) ,\mathcal{E}_{g}\left( X,\eta \right) \right) _{L%
\mathfrak{d}}=\left( \mathcal{E}_{g}\left( Y,\xi \right) ,\left( X,\eta
\right) \right) _{L\mathfrak{d}}
\end{equation*}%
\textit{and} \emph{idempotent} 
\begin{equation*}
\mathcal{E}_{g}\circ \mathcal{E}_{g}=Id_{L\mathfrak{d}}
\end{equation*}%
\textit{It gives rise to a decomposition of} $\mathfrak{g}$ \textit{in two
eigenvalue subspaces}%
\begin{equation*}
\mathcal{E}^{\pm }\left( g\right) =\left\{ X\in \mathfrak{g~}/\mathfrak{~}%
\mathcal{E}_{g}X=\pm X\right\}
\end{equation*}%
\textit{such that} $\mathfrak{g}=\mathcal{E}^{+}\oplus \mathcal{E}^{-}$%
\textit{. Therefore, if }$X$ \textit{is in }$\mathcal{E}^{\pm }\left(
g\right) $\textit{\ it can be written as} 
\begin{equation*}
X=\left( X_{+},\left( \mathcal{B}_{g}\pm \mathcal{G}_{g}\right) X_{+}\right)
\end{equation*}%
\textit{or, alternatively, as} 
\begin{equation*}
X=\left( \left( \mathcal{B}_{g}\pm \mathcal{G}_{g}\right)
^{-1}X_{-},X_{-}\right)
\end{equation*}%
\textit{Observe that if} $X\in \mathcal{E}^{\pm }\left( g\right) $\textit{,
then} $Ad_{g_{+}}^{G}X\in \mathcal{E}^{\pm }\left( e\right) $.
\end{description}

Coming back to the Hamilton function on $G\times \mathfrak{g}^{\ast }$, we
write 
\begin{equation*}
\mathcal{H}\left( g,\eta \right) =\frac{1}{2}\left( \bar{\psi}\left( \mathrm{%
Ad}_{g^{-1}}^{G\ast }\eta +C\left( g\right) \right) ,\mathcal{E}\bar{\psi}%
\left( \mathrm{Ad}_{g^{-1}}^{G\ast }\eta +C\left( g\right) \right) \right) _{%
\mathfrak{g}}
\end{equation*}%
and its differential $d\mathcal{H}=\left( \mathbf{d}\mathcal{H},\delta 
\mathcal{H}\right) $ is%
\begin{equation}
\left\{ 
\begin{array}{l}
\mathbf{d}\mathcal{H}=g^{-1}\psi \left[ \bar{\psi}\left( \eta -C\left(
g^{-1}\right) \right) ,\mathcal{E}_{g}\bar{\psi}\left( \eta -C\left(
g^{-1}\right) \right) \right] \\ 
\qquad ~~~+g^{-1}\mathrm{Ad}_{g}^{G\ast }\hat{c}\left( \mathcal{E}\mathrm{Ad}%
_{g}^{G}\bar{\psi}\left( \eta -C\left( g^{-1}\right) \right) \right) \\ 
\\ 
\delta \mathcal{H}=\mathcal{E}_{g}\bar{\psi}\left( \eta -C\left(
g^{-1}\right) \right)%
\end{array}%
\right.  \label{dif H}
\end{equation}

The Hamilton equations in $G\times \mathfrak{g}^{\ast }$ are be obtained
from the hamiltonian vector field $\left( \ref{ham vect cent exten sympl
form}\right) $ giving%
\begin{equation*}
\left\{ 
\begin{array}{l}
g^{-1}\dot{g}=\mathcal{E}_{g}\bar{\psi}\left( \eta -C\left( g^{-1}\right)
\right) \\ 
\\ 
\dot{\eta}=\mathrm{ad}_{\mathcal{E}_{g}\bar{\psi}\left( \eta -C\left(
g^{-1}\right) \right) }^{\mathfrak{g}\ast }C\left( g^{-1}\right)%
\end{array}%
\right.
\end{equation*}%
It is worth to remark here that, in retrieving the Lagrange equation, we get
the second order equation 
\begin{equation*}
\frac{\partial }{\partial t}\left( \mathcal{E}_{g}\left( g^{-1}\dot{g}%
\right) +\bar{\psi}\left( C\left( g^{-1}\right) \right) \right) =\left[
g^{-1}\dot{g},\bar{\psi}\left( C\left( g^{-1}\right) \right) \right]
\end{equation*}%
where the Lie algebra bracket in the rhs is a manifestation of the
topological WZ term.

\subsection{The Hamilton equation in $\mathcal{N}\left( g_{-},\protect\eta %
_{-}\right) $}

The Hamilton equation in $\mathcal{N}\left( g_{-},\eta _{-}\right) $ where
given in $\left( \ref{Dirac Ham equations}\right) $. Since we shall be
concerned with loop groups, from now on we assume that the cocycle $C$
restricts to the factors as the map $C:G_{\pm }\rightarrow \mathfrak{g}_{\mp
}^{\ast }$.

Then, replacing $\mathbf{d}\mathcal{H},\delta \mathcal{H}$ in those Hamilton
equation by the ones given in $\left( \ref{dif H}\right) $ we get the
Hamilton equations for the Hamilton function $\left( \ref{w2}\right) $.
Before to write them in terms of the components $G_{\pm }\times \mathfrak{g}%
_{\pm }^{\ast }$, we verify that they are indeed generated by an
infinitesimal generator of the action $\left( \ref{SymplecticInducedAction}%
\right) $. In fact, after replacing the differential $\left( \ref{dif H}%
\right) $ in the equations $\left( \ref{Dirac Ham equations}\right) $, we
can handle them to the form%
\begin{equation*}
\left\{ 
\begin{array}{l}
g^{-1}\dot{g}=\mathrm{Ad}_{g_{-}^{-1}}^{G}\Pi _{\mathfrak{g}_{+}}\mathrm{Ad}%
_{g_{+}^{-1}}^{G}\left( \mathrm{Ad}_{g}^{G}\mathcal{E}_{g}\bar{\psi}\left(
\eta -C\left( g^{-1}\right) \right) \right) \\ 
\\ 
\dot{\eta}=-\mathrm{Ad}_{g_{-}}^{G\ast }\Pi _{\mathfrak{g}_{+}^{\ast }}%
\mathrm{Ad}_{g_{-}^{-1}}^{G\ast }\left( \mathrm{ad}_{\mathrm{Ad}%
_{g_{-}^{-1}}^{G}\Pi _{\mathfrak{g}_{-}}\mathrm{Ad}_{g_{+}^{-1}}^{G}\left( 
\mathrm{Ad}_{g}^{G}\mathcal{E}_{g}\bar{\psi}\left( \eta -C\left(
g^{-1}\right) \right) \right) }^{\mathfrak{g}\ast }\eta \right) \\ 
\qquad -\mathrm{Ad}_{g_{-}}^{G\ast }\Pi _{\mathfrak{g}_{+}^{\ast }}\mathrm{Ad%
}_{g_{-}^{-1}}^{G\ast }\left( \hat{c}\left( \mathrm{Ad}_{g^{-1}}^{G}\left( 
\mathrm{Ad}_{g}^{G}\mathcal{E}_{g}\bar{\psi}\left( \eta -C\left(
g^{-1}\right) \right) \right) \right) \right)%
\end{array}%
\right.
\end{equation*}%
verifying that 
\begin{equation*}
\left( \dot{g},\dot{\eta}\right) =V_{\left( \mathrm{Ad}_{g}^{G}\mathcal{E}%
_{g}\bar{\psi}\left( \eta -C\left( g^{-1}\right) \right) ,a\right) }^{%
\mathcal{N}}\left( g,\eta \right)
\end{equation*}%
for $V_{\left( X,a\right) }^{\mathcal{N}}$ defined in $\left( \ref{L7}%
\right) $, as we expected from the collective character of the dynamics in $%
\mathcal{N}\left( g_{-},\eta _{-}\right) $.

Now, let us work out the dynamics in $\mathcal{N}\left( g_{-},\eta
_{-}\right) $ in terms of its components $G_{\pm }\times \mathfrak{g}_{\pm
}^{\ast }$ by writing $g=g_{+}g_{-}$ and $\eta =\eta _{+}+\eta _{-}$. Of
course, $\left( g_{-},\eta _{-}\right) $ remains frozen, while $\left(
g_{+},\eta _{+}\right) $ evolves in $\mathcal{N}\left( g_{-},\eta
_{-}\right) $ ruled by the equations: 
\begin{equation*}
\left\{ 
\begin{array}{l}
g_{+}^{-1}\dot{g}_{+}=\Pi _{\mathfrak{g}_{+}}\mathrm{Ad}_{g_{+}^{-1}}^{G}%
\mathcal{E}\mathrm{Ad}_{g}^{G}\bar{\psi}\left( \eta -C\left( g^{-1}\right)
\right) \\ 
\\ 
\dot{\eta}=-\mathrm{Ad}_{g_{-}}^{G\ast }\Pi _{\mathfrak{g}_{+}^{\ast }}\psi %
\left[ \mathrm{Ad}_{g_{-}}^{G}\bar{\psi}\left( \eta -C\left( g^{-1}\right)
\right) ,\Pi _{\mathfrak{g}_{-}}\mathrm{Ad}_{g_{-}}^{G}\mathcal{E}_{g}\bar{%
\psi}\left( \eta -C\left( g^{-1}\right) \right) \right] \\ 
~~~~~~~+\mathrm{Ad}_{g_{-}}^{G\ast }\Pi _{\mathfrak{g}_{+}^{\ast }}\psi %
\left[ \mathrm{Ad}_{g_{-}}^{G}\bar{\psi}\left( C\left( g^{-1}\right) \right)
,\Pi _{\mathfrak{g}_{+}}\mathrm{Ad}_{g_{-}}^{G}\mathcal{E}_{g}\bar{\psi}%
\left( \eta -C\left( g^{-1}\right) \right) \right] \\ 
~~~~~~~-\mathrm{Ad}_{g_{-}}^{G\ast }\Pi _{\mathfrak{g}_{+}^{\ast }}\mathrm{Ad%
}_{g_{+}}^{G\ast }\hat{c}\left( \mathcal{E}\mathrm{Ad}_{g}^{G}\bar{\psi}%
\left( \eta -C\left( g^{-1}\right) \right) \right)%
\end{array}%
\right.
\end{equation*}%
Here we observe that, by taking $\left( g_{-},\eta _{-}\right) \in \ker
C\times \mathrm{Char}\left( \mathfrak{g}_{-}\right) $ it happens that 
\begin{equation*}
C\left( g^{-1}\right) =C\left( g_{-}^{-1}g_{+}^{-1}\right) =\mathrm{Ad}%
_{g_{-}}^{G\ast }C\left( g_{+}^{-1}\right)
\end{equation*}%
Then, having since for $\eta _{\pm }\in \mathfrak{g}_{\pm }^{\ast }$ and $%
g_{\mp }\in G_{\mp }$, $\mathrm{Ad}_{g_{\mp }}^{G\ast }\eta _{\pm }$ $\in $%
\textit{\ }$\mathfrak{g}_{\pm }^{\ast }$, which also means that\textit{\ }$%
\mathrm{ad}_{X_{\mp }}^{\mathfrak{g}\ast }\eta _{\pm }=\Pi _{\mathfrak{g}%
_{+}^{\ast }}\mathrm{ad}_{X_{\mp }}^{\mathfrak{g}\ast }\eta _{\pm }$. Again,
using these facts and the properties of the different cocycles, we arrive to%
\begin{equation*}
\left\{ 
\begin{array}{l}
g_{+}^{-1}\dot{g}_{+}=\Pi _{\mathfrak{g}_{+}}\mathcal{E}_{g_{+}}\bar{\psi}%
\left( \mathrm{Ad}_{g_{-}^{-1}}^{G\ast }\eta -C\left( g_{+}^{-1}\right)
\right) \\ 
\\ 
\mathrm{Ad}_{g_{-}^{-1}}^{G\ast }\dot{\eta}=-\Pi _{\mathfrak{g}_{+}^{\ast
}}\psi \left[ \bar{\psi}\left( \mathrm{Ad}_{g_{-}^{-1}}^{G\ast }\eta \right)
,\Pi _{\mathfrak{g}_{-}}\mathcal{E}_{g_{+}}\bar{\psi}\left( \mathrm{Ad}%
_{g_{-}^{-1}}^{G\ast }\eta -C\left( g_{+}^{-1}\right) \right) \right] \\ 
~~~~~~~~~~~~~~~-\Pi _{\mathfrak{g}_{+}^{\ast }}\hat{c}\left( \Pi _{\mathfrak{%
g}_{-}}\mathcal{E}_{g_{+}}\bar{\psi}\left( \mathrm{Ad}_{g_{-}^{-1}}^{G\ast
}\eta -C\left( g_{+}^{-1}\right) \right) \right)%
\end{array}%
\right.
\end{equation*}

\section{The Lagrange equation}

We now proceed to retrieve the Lagrange equation on $\mathcal{N}\left(
g_{-},\eta _{-}\right) $ by replacing the first Hamilton equation into the
second one. Therefore, since $\eta _{-}$ is a character of $\mathfrak{g}_{-}$%
,%
\begin{equation*}
\Pi _{\mathfrak{g}_{-}^{\ast }}\mathrm{Ad}_{g_{-}^{-1}}^{G\ast }\eta
_{-}=Ad_{g_{-}^{-1}}^{\ast }\eta _{-}=\eta _{-}
\end{equation*}%
the first Hamilton equation becomes in%
\begin{equation*}
g_{+}^{-1}\dot{g}_{+}=\mathcal{G}_{g_{+}}^{-1}\bar{\psi}\left( \mathrm{Ad}%
_{g_{-}^{-1}}^{G\ast }\eta _{+}\right) +\mathcal{G}_{g_{+}}^{-1}\bar{\psi}%
\left( \Pi _{\mathfrak{g}_{+}^{\ast }}\mathrm{Ad}_{g_{-}^{-1}}^{G\ast }\eta
_{-}\right) +\mathcal{G}_{g_{+}}^{-1}\mathcal{B}_{g_{+}}\bar{\psi}\left(
C\left( g_{+}^{-1}\right) -\eta _{-}\right)
\end{equation*}%
then 
\begin{equation}
\bar{\psi}\left( \mathrm{Ad}_{g_{-}^{-1}}^{G\ast }\eta _{+}\right) =\mathcal{%
G}_{g_{+}}g_{+}^{-1}\dot{g}_{+}-\mathcal{B}_{g_{+}}\bar{\psi}\left( C\left(
g_{+}^{-1}\right) \right) +\left( \mathcal{B}_{g_{+}}-\Pi _{\mathfrak{g}_{-}}%
\mathrm{Ad}_{g_{-}}^{G}\right) \bar{\psi}\left( \eta _{-}\right)
\label{eta 2 gdot}
\end{equation}

The second Hamilton equation can be handled to 
\begin{eqnarray*}
&&\bar{\psi}\left( Ad_{g_{-}^{-1}}^{G\ast }\dot{\eta}\right) \\
&=&-\Pi _{\mathfrak{g}_{-}}\left[ \bar{\psi}\left( \mathrm{Ad}%
_{g_{-}^{-1}}^{G\ast }\eta \right) ,\left( \mathcal{G}_{g_{+}}-\mathcal{B}%
_{g_{+}}(\mathcal{G}_{g_{+}})^{-1}\mathcal{B}_{g_{+}}\right) \Pi _{\mathfrak{%
g}_{+}}\bar{\psi}\left( \mathrm{Ad}_{g_{-}^{-1}}^{G\ast }\eta -C\left(
g_{+}^{-1}\right) \right) \right] \\
&&-\Pi _{\mathfrak{g}_{-}}\left[ \bar{\psi}\left( \mathrm{Ad}%
_{g_{-}^{-1}}^{G\ast }\eta \right) ,\mathcal{B}_{g_{+}}(\mathcal{G}%
_{g_{+}})^{-1}\Pi _{\mathfrak{g}_{-}}\bar{\psi}\left( \mathrm{Ad}%
_{g_{-}^{-1}}^{G\ast }\eta -C\left( g_{+}^{-1}\right) \right) \right] \\
&&-\bar{\psi}\left( \hat{c}\left( \left( \mathcal{G}_{g_{+}}-\mathcal{B}%
_{g_{+}}(\mathcal{G}_{g_{+}})^{-1}\mathcal{B}_{g_{+}}\right) \Pi _{\mathfrak{%
g}_{+}}\bar{\psi}\left( \mathrm{Ad}_{g_{-}^{-1}}^{G\ast }\eta -C\left(
g_{+}^{-1}\right) \right) \right) \right) \\
&&-\bar{\psi}\left( \hat{c}\left( \mathcal{B}_{g_{+}}(\mathcal{G}%
_{g_{+}})^{-1}\Pi _{\mathfrak{g}_{-}}\bar{\psi}\left( \mathrm{Ad}%
_{g_{-}^{-1}}^{G\ast }\eta -C\left( g_{+}^{-1}\right) \right) \right) \right)
\end{eqnarray*}%
and by substituting $\bar{\psi}\left( Ad_{g_{-}^{-1}}^{G\ast }\eta
_{+}\right) $ as given in eq. $\left( \ref{eta 2 gdot}\right) $ we get%
\begin{equation}
\begin{array}{l}
\dfrac{d}{dt}\left( \mathcal{G}_{g_{+}}g_{+}^{-1}\dot{g}_{+}-\mathcal{B}%
_{g_{+}}\bar{\psi}\left( C\left( g_{+}^{-1}\right) \right) +\mathcal{B}%
_{g_{+}}\bar{\psi}\left( \eta _{-}\right) \right) \\ 
=-\Pi _{\mathfrak{g}_{-}}\left[ \mathcal{G}_{g_{+}}g_{+}^{-1}\dot{g}_{+}-%
\mathcal{B}_{g_{+}}\bar{\psi}\left( C\left( g_{+}^{-1}-\eta _{-}\right)
\right) ,\mathcal{B}_{g_{+}}g_{+}^{-1}\dot{g}_{+}-\mathcal{G}_{g_{+}}\bar{%
\psi}\left( C\left( g_{+}^{-1}\right) -\eta _{-}\right) \right] \\ 
~~~-\Pi _{\mathfrak{g}_{-}}\left[ \bar{\psi}\left( \eta _{-}\right) ,%
\mathcal{B}_{g_{+}}g_{+}^{-1}\dot{g}_{+}-\mathcal{G}_{g_{+}}\bar{\psi}\left(
C\left( g_{+}^{-1}\right) -\eta _{-}\right) \right] \\ 
~~~-\bar{\psi}\left( \hat{c}\left( \mathcal{B}_{g_{+}}g_{+}^{-1}\dot{g}_{+}-%
\mathcal{G}_{g_{+}}\bar{\psi}\left( C\left( g_{+}^{-1}\right) -\eta
_{-}\right) \right) \right)%
\end{array}
\label{Lagrange eq}
\end{equation}

On the fiber $\mathcal{N}\left( e,0\right) \cong G_{+}\times \mathfrak{g}%
_{+}^{\ast }$ it reduces to 
\begin{eqnarray*}
&&\dfrac{d}{dt}\psi \left( \mathcal{G}_{g_{+}}g_{+}^{-1}\dot{g}_{+}-\mathcal{%
B}_{g_{+}}\bar{\psi}\left( C\left( g_{+}^{-1}\right) \right) \right) +\hat{c}%
\left( \mathcal{B}_{g_{+}}g_{+}^{-1}\dot{g}_{+}-\mathcal{G}_{g_{+}}\bar{\psi}%
\left( C\left( g_{+}^{-1}\right) \right) \right) \\
&=&-\psi \left[ \mathcal{G}_{g_{+}}g_{+}^{-1}\dot{g}_{+}-\mathcal{B}_{g_{+}}%
\bar{\psi}\left( C\left( g_{+}^{-1}\right) \right) ,\mathcal{B}%
_{g_{+}}g_{+}^{-1}\dot{g}_{+}-\mathcal{G}_{g_{+}}\bar{\psi}\left( C\left(
g_{+}^{-1}\right) \right) \right]
\end{eqnarray*}

\section{The Lagrangian function on $\mathcal{N}\left( g_{-},\protect\eta %
_{-}\right) $}

Now, we built up the Lagrangian version on the fibers $\mathcal{N}\left(
g_{-},\eta _{-}\right) $, which in turn are symplectic submanifolds equipped
with the restriction of the symplectic form on $G\times \mathfrak{g}^{\ast }$%
. The Lagrangian relates with the Hamilton function by 
\begin{equation*}
dL=\Theta -dH
\end{equation*}%
provided there exist the $1$-form $\Theta $ such satisfying 
\begin{equation*}
\left. \omega _{c}\right\vert \mathcal{N}\left( g_{-},\eta _{-}\right)
=-d\Theta
\end{equation*}%
In the current case, it is supplied by the following proposition.

\begin{description}
\item[Proposition:] \textit{The restriction of the symplectic form }$\omega
_{c}$ \textit{to} $\mathcal{N}\left( g_{-},\eta _{-}\right) $ \textit{is an
exact form such that} 
\begin{equation*}
\left. \omega _{c}\right\vert \mathcal{N}\left( g_{-},\eta _{-}\right)
=-d\Theta
\end{equation*}%
\textit{with }$\Theta $ \textit{being the restriction of the canonical }$1$%
\textit{-form of }$G\times \mathfrak{g}^{\ast }$ \textit{to} $\mathcal{N}%
\left( g_{-},\eta _{-}\right) $ 
\begin{equation*}
\left\langle \Theta ,\left( g_{+}\left( \bar{\psi}\left(
Ad_{g_{-}^{-1}}^{\ast }\psi \left( X_{+}\right) \right) \right) g_{-},\xi
_{+}\right) \right\rangle _{\left( g,\eta \right) }:=\left\langle \eta
,g_{+}\left( \bar{\psi}\left( Ad_{g_{-}^{-1}}^{\ast }\psi \left(
X_{+}\right) \right) \right) g_{-}\right\rangle
\end{equation*}
\end{description}

\textbf{Proof: }By the isotropic character of the subspaces $\mathfrak{g}%
_{\pm }$ under the cocycle $c$, the restriction of $\omega _{c}$ reduces to 
\begin{eqnarray*}
&&\left\langle \omega ,\left( g_{+}\left( \bar{\psi}\left(
Ad_{g_{-}^{-1}}^{\ast }\psi \left( X_{+}\right) \right) \right) g_{-},\xi
_{+}\right) \otimes \left( g_{+}\left( \bar{\psi}\left(
Ad_{g_{-}^{-1}}^{\ast }\psi \left( Y_{+}\right) \right) \right)
g_{-},\lambda _{+}\right) \right\rangle _{\left( g,\eta \right) } \\
&=&-\left\langle \xi _{+},\mathrm{Ad}_{g_{-}^{-1}}^{G}\left( \bar{\psi}%
\left( Ad_{g_{-}^{-1}}^{\ast }\psi \left( Y_{+}\right) \right) \right)
\right\rangle +\left\langle \lambda _{+},\mathrm{Ad}_{g_{-}^{-1}}^{G}\left( 
\bar{\psi}\left( Ad_{g_{-}^{-1}}^{\ast }\psi \left( X_{+}\right) \right)
\right) \right\rangle \\
&&+\left\langle \eta ,\left[ \mathrm{Ad}_{g_{-}^{-1}}^{G}\left( \bar{\psi}%
\left( Ad_{g_{-}^{-1}}^{\ast }\psi \left( X_{+}\right) \right) \right) ,%
\mathrm{Ad}_{g_{-}^{-1}}^{G}\left( \bar{\psi}\left( Ad_{g_{-}^{-1}}^{\ast
}\psi \left( Y_{+}\right) \right) \right) \right] \right\rangle
\end{eqnarray*}%
Let us define it as $\Theta \in T^{\ast }\mathcal{N}\left( g_{-},\eta
_{-}\right) $ such that 
\begin{equation*}
\left\langle \Theta ,\left( g_{+}\left( \bar{\psi}\left(
Ad_{g_{-}^{-1}}^{\ast }\psi \left( X_{+}\right) \right) \right) g_{-},\xi
_{+}\right) \right\rangle _{\left( g,\eta \right) }:=\left\langle \eta
,g_{+}\left( \bar{\psi}\left( Ad_{g_{-}^{-1}}^{\ast }\psi \left(
X_{+}\right) \right) \right) g_{-}\right\rangle
\end{equation*}%
then 
\begin{eqnarray*}
&&\left\langle d\Theta ,\left( g_{+}\left( \bar{\psi}\left(
Ad_{g_{-}^{-1}}^{\ast }\psi \left( X_{+}\right) \right) \right) g_{-},\xi
_{+}\right) \otimes \left( g_{+}\left( \bar{\psi}\left(
Ad_{g_{-}^{-1}}^{\ast }\psi \left( Y_{+}\right) \right) \right)
g_{-},\lambda _{+}\right) \right\rangle _{\left( g,\eta \right) } \\
&=&\left\langle \xi _{+},g_{+}\left( \bar{\psi}\left( Ad_{g_{-}^{-1}}^{\ast
}\psi \left( Y_{+}\right) \right) \right) g_{-}\right\rangle -\left\langle
\lambda _{+},g_{+}\left( \bar{\psi}\left( Ad_{g_{-}^{-1}}^{\ast }\psi \left(
X_{+}\right) \right) \right) g_{-}\right\rangle \\
&&-\left\langle \eta ,\left[ g_{+}\left( \bar{\psi}\left(
Ad_{g_{-}^{-1}}^{\ast }\psi \left( X_{+}\right) \right) \right)
g_{-},g_{+}\left( \bar{\psi}\left( Ad_{g_{-}^{-1}}^{\ast }\psi \left(
Y_{+}\right) \right) \right) g_{-}\right] \right\rangle
\end{eqnarray*}%
showing that $\left. \omega _{c}\right\vert \mathcal{N}\left( g_{-},\eta
_{-}\right) =-d\Theta $ as stated.$\blacksquare $

Therefore, the Lagrangian function on $\mathcal{N}\left( g_{-},\eta
_{-}\right) $ is 
\begin{equation*}
L_{\mathcal{N}\left( g_{-},\eta _{-}\right) }\left( g,\dot{g}\right)
=\left\langle \eta ,g^{-1}\dot{g}\right\rangle -\left. \mathcal{H}\left(
g,\eta \right) \right\vert _{\mathcal{N}\left( g_{-},\eta _{-}\right) }
\end{equation*}%
where $\left. \mathcal{H}\left( g,\eta \right) \right\vert _{\mathcal{N}%
\left( g_{-},\eta _{-}\right) }$ is the restriction of $\mathcal{H}\left(
g,\eta \right) $ to $\mathcal{N}\left( g_{-},\eta _{-}\right) $, with 
\begin{equation*}
\mathcal{H}\left( g,\eta \right) =\frac{1}{2}\left( \bar{\psi}\left(
J_{B}^{L\wedge }\left( g,\eta \right) \right) ,\mathcal{E}\bar{\psi}\left(
J_{B}^{L\wedge }\left( g,\eta \right) \right) \right) _{\mathfrak{g}^{\wedge
}}
\end{equation*}%
where, having in mind that $\left( g_{-},\eta _{-}\right) \in \ker C\times 
\mathrm{Char}\left( \mathfrak{g}_{-}\right) $, 
\begin{equation*}
\left. J_{B}^{L\wedge }(g,\eta )\right\vert _{\mathcal{N}\left( g_{-},\eta
_{-}\right) }=\left( \mathrm{Ad}_{g^{-1}}^{G\ast }\eta +C\left( g_{+}\right)
,\,1\right)
\end{equation*}%
Then%
\begin{eqnarray}
H(g_{+},\eta _{+}) &:&=\left. \mathcal{H}\left( g,\eta \right) \right\vert _{%
\mathcal{N}\left( g_{-},\eta _{-}\right) }  \label{LN2} \\
&=&\frac{1}{2}\left( \bar{\psi}\left( \mathrm{Ad}_{g^{-1}}^{G\ast }\eta
+C\left( g_{+}\right) \right) ,\mathcal{E}\bar{\psi}\left( \mathrm{Ad}%
_{g^{-1}}^{G\ast }\eta +C\left( g_{+}\right) \right) \right) _{\mathfrak{g}%
^{\wedge }}  \notag
\end{eqnarray}

On the other side, on $\mathcal{N}\left( g_{-},\eta _{-}\right) $ we have
that 
\begin{equation*}
g^{-1}\dot{g}=g_{-}^{-1}g_{+}^{-1}\dot{g}_{+}g_{-}=Ad_{g_{-}^{-1}}^{G}\left(
g_{+}^{-1}\dot{g}_{+}\right)
\end{equation*}%
therefore 
\begin{eqnarray*}
&&L_{\mathcal{N}}\left( g,\dot{g}\right) \\
&=&\left\langle \eta ,\mathrm{Ad}_{g_{-}^{-1}}^{G}\left( g_{+}^{-1}\dot{g}%
_{+}\right) \right\rangle -\frac{1}{2}\left( \bar{\psi}\left( \mathrm{Ad}%
_{g^{-1}}^{G\ast }\eta +C\left( g_{+}\right) \right) ,\mathcal{E}\bar{\psi}%
\left( \mathrm{Ad}_{g^{-1}}^{G\ast }\eta +C\left( g_{+}\right) \right)
\right) _{\mathfrak{g}}
\end{eqnarray*}%
that is equivalent to 
\begin{eqnarray*}
&&L_{\mathcal{N}}\left( g,\dot{g}\right) \\
&=&\left( \bar{\psi}\left( \mathrm{Ad}_{g_{-}^{-1}}^{G\ast }\eta \right)
,g_{+}^{-1}\dot{g}_{+}\right) _{\mathfrak{g}}-\frac{1}{2}\left( \bar{\psi}%
\left( \mathrm{Ad}_{g_{-}^{-1}}^{G\ast }\eta \right) ,\mathcal{E}_{g_{+}}%
\bar{\psi}\left( \mathrm{Ad}_{g_{-}^{-1}}^{G\ast }\eta \right) \right) _{%
\mathfrak{g}} \\
&&+\left( \bar{\psi}\left( \mathrm{Ad}_{g_{-}^{-1}}^{G\ast }\eta \right) ,%
\mathcal{E}_{g_{+}}\bar{\psi}\left( C\left( g_{+}^{-1}\right) \right)
\right) _{\mathfrak{g}}-\frac{1}{2}\left( \bar{\psi}\left( C\left(
g_{+}^{-1}\right) \right) ,\mathcal{E}_{g_{+}}\bar{\psi}\left( C\left(
g_{+}^{-1}\right) \right) \right) _{\mathfrak{g}}
\end{eqnarray*}

By replacing the fiber coordinate $\eta _{+}$ as a function of the velocity,
as it was obtained from the first Hamilton equation in eq. $\left( \ref{eta
2 gdot}\right) $, 
\begin{equation*}
Ad_{g_{-}^{-1}}^{G\ast }\eta _{+}=\psi \left( \mathcal{G}_{g_{+}}g_{+}^{-1}%
\dot{g}_{+}-\mathcal{B}_{g_{+}}\bar{\psi}\left( C\left( g_{+}^{-1}\right)
\right) +\left( \mathcal{B}_{g_{+}}-\Pi _{\mathfrak{g}_{-}}Ad_{g_{-}}^{G}%
\right) \bar{\psi}\left( \eta _{-}\right) \right)
\end{equation*}%
we get%
\begin{eqnarray}
L_{\mathcal{N}}\left( g,\dot{g}\right) &=&\frac{1}{2}\left( \mathcal{G}%
_{g_{+}}g_{+}^{-1}\dot{g}_{+},g_{+}^{-1}\dot{g}_{+}\right) _{\mathfrak{g}%
}-\left( g_{+}^{-1}\dot{g}_{+},\mathcal{B}_{g}\bar{\psi}\left( C\left(
g_{+}^{-1}\right) -\eta _{-}\right) \right) _{\mathfrak{g}}  \label{LN5} \\
&&-\frac{1}{2}\left( \bar{\psi}\left( C\left( g_{+}^{-1}\right) -\eta
_{-}\right) ,\mathcal{G}_{g}\bar{\psi}\left( C\left( g_{+}^{-1}\right) -\eta
_{-}\right) \right) _{\mathfrak{g}}  \notag
\end{eqnarray}%
Finally, it can be written also as the Poisson-Lie $\sigma $-model:

\begin{equation}
L_{\mathcal{N}}\left( g,\dot{g}\right) =\frac{1}{2}\left( \mathcal{R}%
_{g_{+}}^{+}\left( g_{+}^{-1}\dot{g}_{+}-\bar{\psi}\left( C\left(
g_{+}^{-1}\right) -\eta _{-}\right) \right) ,\left( g_{+}^{-1}\dot{g}_{+}+%
\bar{\psi}\left( C\left( g_{+}^{-1}\right) -\eta _{-}\right) \right) \right)
_{\mathfrak{g}}  \notag
\end{equation}%
where $\mathcal{R}_{g_{+}}^{\pm }:=\mathcal{B}_{g_{+}}\pm \mathcal{G}%
_{g_{+}} $.

Thus, we have shown that the collective hamiltonian system $\left( \ref{LN2}%
\right) $ is the phase space version of the lagrangian system introduced in
the context Poisson-Lie T-duality by \cite{KS-1}. Amazingly, it arises
through the Dirac restriction method of the larger phase $G\times \mathfrak{g%
}^{\ast }$ of a noncollective nonAd-invariant hamiltonian function.

\subsection{Centrally extended loop algebras}

Let consider $\mathfrak{g}$ as being the loop algebra $\mathfrak{g}=\mathrm{L%
}\mathfrak{h}$ for some Lie algebra $\mathfrak{h}$ equipped with the
bilinear form 
\begin{equation*}
\left( X,Y\right) _{\mathfrak{g}}=\frac{1}{2\pi }\int_{S^{1}}\left( X\left(
s\right) ,Y\left( s\right) \right) _{\mathfrak{h}}\,ds
\end{equation*}%
Then we define the coadjoint cocycle $C_{\mathrm{k}}:H\rightarrow \mathfrak{h%
}^{\ast }$, 
\begin{equation}
C_{\mathrm{k}}\left( g\right) =\mathrm{k}\psi \left( g^{\prime }g^{-1}\right)
\label{cc2}
\end{equation}%
which satisfy%
\begin{equation*}
C_{\mathrm{k}}\left( gh\right) =\mathrm{Ad}_{g^{-1}}^{G\ast }C_{\mathrm{k}%
}\left( h\right) +C_{\mathrm{k}}\left( g\right)
\end{equation*}%
It extend to a cocycle on $G=LH$ by the identification of the dual $%
\mathfrak{g}^{\ast }$ through the bilinear form above defined, in such a way
that%
\begin{equation*}
\left\langle C_{\mathrm{k}}\left( g\right) ,Y\right\rangle =\frac{\mathrm{k}%
}{2\pi }\int_{S^{1}}\left( g^{\prime }g^{-1},Y\right) _{\mathfrak{h}}ds
\end{equation*}%
so we define $c_{\mathrm{k}}:\mathfrak{g}\otimes \mathfrak{g}\longrightarrow 
\mathbb{R}$ as 
\begin{equation*}
c_{\mathrm{k}}(X,Y)=\frac{\mathrm{k}}{2\pi }\int_{S^{1}}\left( X\left(
s\right) ,Y^{\prime }\left( s\right) \right) _{\mathfrak{h}}\,ds
\end{equation*}%
with the map $\hat{c}_{\mathrm{k}}:\mathfrak{g}\longrightarrow \mathfrak{g}%
^{\ast }$ 
\begin{equation}
\hat{c}_{\mathrm{k}}(X)=-\mathrm{k}\psi \left( X^{\prime }\right)
\label{cc1}
\end{equation}

Let us substitute these cocycle $\left( \ref{cc2}\right) $ and $\left( \ref%
{cc1}\right) $ in the Lagrange equation $\left( \ref{Lagrange eq}\right) $,
to obtain 
\begin{equation*}
\begin{array}{l}
\dfrac{\partial }{\partial t}\left( \mathcal{G}_{g_{+}}g_{+}^{-1}\dot{g}_{+}+%
\mathrm{k}\mathcal{B}_{g_{+}}g_{+}^{-1}g_{+}^{\prime }+\mathcal{B}_{g_{+}}%
\bar{\psi}\left( \eta _{-}\right) +\bar{\psi}\left( \eta _{-}\right) \right)
\\ 
-\mathrm{k}\dfrac{\partial }{\partial x}\left( \mathcal{B}_{g_{+}}g_{+}^{-1}%
\dot{g}_{+}+\mathrm{k}\mathcal{G}_{g_{+}}g_{+}^{-1}g_{+}^{\prime }+\mathcal{G%
}_{g_{+}}\bar{\psi}\left( \eta _{-}\right) \right) \\ 
=\Pi _{\mathfrak{g}_{-}}\left[ \mathcal{B}_{g_{+}}g_{+}^{-1}\dot{g}_{+}+%
\mathcal{G}_{g_{+}}\left( \mathrm{k}g_{+}^{-1}g_{+}^{\prime }+\bar{\psi}%
\left( \eta _{-}\right) \right) ,\mathcal{G}_{g_{+}}g_{+}^{-1}\dot{g}_{+}+%
\mathcal{B}_{g_{+}}\left( \mathrm{k}g_{+}^{-1}g_{+}^{\prime }+\bar{\psi}%
\left( \eta _{-}\right) \right) +\bar{\psi}\left( \eta _{-}\right) \right]%
\end{array}%
\end{equation*}%
In this case, and introducing $\partial _{\pm }=\dfrac{\partial }{\partial t}%
\pm \mathrm{k}\dfrac{\partial }{\partial x}$, the Lagrangian density $\left( %
\ref{LN5}\right) $ turns in%
\begin{equation}
\mathcal{L}_{\mathcal{N}}\left( g_{+}\right) =\frac{1}{2}\left( \mathcal{R}%
_{g_{+}}\left( g_{+}^{-1}\partial _{+}g_{+}+\bar{\psi}\left( \eta
_{-}\right) \right) ,\left( g_{+}^{-1}\partial _{-}g_{+}-\bar{\psi}\left(
\eta _{-}\right) \right) \right) _{\mathfrak{g}}  \label{Lagrangian PS1}
\end{equation}

In order to eliminate the $g_{+}$-dependence in the operator $\mathcal{R}%
_{g_{+}}$, leaving a purely world sheet depending one, we use the
decomposition of the Lie algebra $\mathfrak{g}$ as the direct sum of the
eingenspaces of the operator $\mathcal{E}_{g_{+}}$ following reference \cite%
{Majid-Begg}. Since $\left( \left( \mathcal{R}_{e}^{\pm }\right)
^{-1}X_{-},X_{-}\right) \in \mathcal{E}^{\pm }\left( e\right) $, $\mathrm{Ad}%
_{g_{+}^{-1}}^{G}\left( \left( \mathcal{R}_{e}^{\pm }\right)
^{-1}X_{-},X_{-}\right) \in \mathcal{E}^{\pm }\left( g\right) $ which
implies the relation%
\begin{equation*}
\mathrm{Ad}_{g_{+}}^{G}\left( \mathcal{R}_{g}^{\pm }\right) ^{-1}\Pi _{%
\mathfrak{g}_{-}}\mathrm{Ad}_{g_{+}^{-1}}^{G}X_{-}=\left( \mathcal{R}%
_{e}^{\pm }\right) ^{-1}X_{-}+\Pi _{\mathfrak{g}_{+}}\mathrm{Ad}%
_{g_{+}}^{G}\Pi _{\mathfrak{g}_{+}}\mathrm{Ad}_{g_{+}^{-1}}^{G}X_{-}
\end{equation*}%
By using that the right translated Poisson-Lie bivector on $G_{+}$, $\pi
_{+}^{R}:G_{+}\longrightarrow \mathfrak{g}_{+}\otimes \mathfrak{g}_{+}$, is
defined from the relation \cite{Lu-We}\textit{\ }%
\begin{equation*}
\left\langle \psi \left( X_{-}^{\prime }\right) \otimes \psi \left(
X_{-}^{\prime \prime }\right) ,\pi _{+}^{R}\left( g_{+}\right) \right\rangle
:=\left( \Pi _{-}\mathrm{Ad}_{g_{+}^{-1}}^{G}X_{-}^{\prime },\Pi _{+}\mathrm{%
Ad}_{g_{+}^{-1}}^{G}X_{-}^{\prime \prime }\right) _{\mathfrak{g}}
\end{equation*}%
and regarding it as the linear map $\pi _{+}^{R}\left( g_{+}\right) :%
\mathfrak{g}_{-}\longrightarrow \mathfrak{g}_{+}$ such that 
\begin{equation*}
\left( \pi _{+}^{R}\left( g_{+}\right) X_{-},Y_{-}\right) _{\mathfrak{g}%
}:=\left\langle \psi \left( Y_{-}\right) \otimes \psi \left( X_{-}\right)
,\pi _{+}^{R}\left( g_{+}\right) \right\rangle
\end{equation*}%
$\forall Y_{-}\in \mathfrak{g}_{-}$, we identify 
\begin{equation*}
\pi _{+}^{R}\left( g_{+}\right) =-\Pi _{\mathfrak{g}_{+}}\mathrm{Ad}%
_{g_{+}}^{G}\Pi _{\mathfrak{g}_{+}}\mathrm{Ad}_{g_{+}^{-1}}^{G}\Pi _{-}
\end{equation*}%
So, the above relation turns in 
\begin{equation*}
\mathrm{Ad}_{g_{+}}^{G}\left( \mathcal{R}_{g}^{\pm }\right) ^{-1}\Pi _{%
\mathfrak{g}_{-}}\mathrm{Ad}_{g_{+}^{-1}}^{G}\Pi _{\mathfrak{g}_{-}}=\left(
\left( \mathcal{R}_{e}^{\pm }\right) ^{-1}-\pi _{+}^{R}\left( g_{+}\right)
\right) \Pi _{\mathfrak{g}_{-}}
\end{equation*}%
which allows to substitute can be substitute $\mathcal{R}_{g}^{\pm }$ into
the Lagrangian density $\left( \ref{Lagrangian PS1}\right) $ to get%
\begin{equation*}
\mathcal{L}_{\mathcal{N}}\left( g\right) =\frac{1}{2}\left( \left( \left( 
\mathcal{R}_{e}^{\pm }\right) ^{-1}-\pi _{+}^{R}\left( g_{+}\right) \right)
^{-1}\left( g_{+}^{-1}\partial _{+}g_{+}+\bar{\psi}\left( \eta _{-}\right)
\right) ,\left( g_{+}^{-1}\partial _{-}g_{+}-\bar{\psi}\left( \eta
_{-}\right) \right) \right) _{\mathfrak{g}}
\end{equation*}%
that coincides with the T-dual sigma model on the target $G_{+}$ introduced
by Klimcik and Severa \cite{KS-1}.

\bigskip \bigskip

{\LARGE \textbf{Conclusions}}

\bigskip

We have studied the restriction to a family of submanifolds of the second
class constraint type in the cotangent bundle of a double Poisson-Lie group,
equipped with a cocycle extended symplectic form. For a given hamiltonian
system on this phase space, it is this extension which give rise to the 
\emph{WZ }topological term in the corresponding action. Then. we build up
the corresponding Dirac brackets showing that, for the usual $2$-cocycle in
loop groups, it exhibit no contribution coming from it. It is in this sense
that we say that a WZNW phase space restricts to a $\sigma $-model one.
Looking for symmetries on the constrained phase spaces, we worked out the
left translation action despite it is manifestly not a symmetry in the whole
space because of the 2-cocycle is defined on right invariant vector fields.
However, we have shown that in the fiber spaces on points $\left( g_{-},\eta
_{-}\right) \in \ker C\times \mathrm{Char}\left( \mathfrak{g}_{-}\right) $,
the left translation momentum function close a Lie algebra under the
corresponding Dirac bracket, so the left translation symmetry becomes
restored.

This facts suggested to studied collective dynamical systems on these phase
subspaces, which in some way resembles the situation in the hamiltonian
approach WZNW model, see ref. \cite{Harnad}, where the Marsden-Weinstein
reduction procedure has a first stage of reduction in relation with the
obvious right translation symmetry. However, there still remains a residual
left translation symmetry which, once reduced, leads to the chiral modes
solutions.

Thus we studied a hamiltonian system on $G\times \mathfrak{g}^{\ast }$ ruled
by a Hamilton function of the momentum maps of left translation in the
constrained submanifolds, regarded as functions on the whole space. This
hamiltonian turns collective when restricted to $\mathcal{N}\left(
g_{-},\eta _{-}\right) $, and leads to nice dynamics flowing along orbits of
some curves in the group $G$ through the action found in $\left( \ref%
{SymplecticInducedAction}\right) $. Amazingly, these restricted hamiltonian
systems turn to be Poisson-Lie $\sigma $-models, whit some additional terms
involving the parameter $\eta _{-}$ of the fiber. In summary, starting from
a WZNW type model, we arrived through a restriction procedure implemented by
the Dirac method to the Poisson-Lie $\sigma $-models.

Most of the issues developed here are much involved with the hamiltonian
framework for the Poisson-Lie T-duality, as it is clear from the restriction
of the left translation symmetry to the special fibers. The T-dual side can
be obtained, with some subtleties, by considering the fibration $G\times 
\mathfrak{g}^{\ast }\longrightarrow G_{+}\times \mathfrak{g}_{+}^{\ast }$ as
starting point, and repeating the above construction.

Although the current work is purely classical, this formulation as a second
class constrained systems allow to address the quantization in the scheme of
functional integral as a constrained system defined on the whole $G\times 
\mathfrak{g}^{\ast }$.

\bigskip \bigskip

{\LARGE \textbf{Acknowledgments}}

\bigskip

H.M. thanks to CONICET, Argentina, for financial support.

\end{document}